%% file: aipsamp.tex
\documentclass[%
 aip,
 amsmath,amssymb,
 reprint,%
author-year,
]{revtex4-1}

\usepackage{graphicx}
\usepackage{dcolumn}
\usepackage{bm}

\usepackage[utf8]{inputenc}
\usepackage[T1]{fontenc}
\usepackage{mathptmx}
\usepackage{etoolbox}
\usepackage{hyperref}
\hypersetup{
    colorlinks=true,
    linkcolor=black,
    filecolor=black,      
    urlcolor=black,
    citecolor=black,
    }
\usepackage{subcaption}
\usepackage{multirow}
\setcitestyle{authoryear}
\usepackage[colorinlistoftodos]{todonotes}
\newcommand{\todURi}[1]%
{\todo[color=blue!10,inline]{{\bf Uli:} #1}}
\newcommand{\todUR}[1]%
{\todo[color=blue!10,linecolor=darkgray]{\footnotesize {\bf Uli:} #1}}

\usepackage{xspace}

\newcommand{\walberla}{\textsc{waLBerla}\xspace}
\newcommand{\nvidia}{NVIDIA\xspace}

\makeatletter
\def\@email#1#2{%
 \endgroup
 \patchcmd{\titleblock@produce}
  {\frontmatter@RRAPformat}
  {\frontmatter@RRAPformat{\produce@RRAP{*#1\href{mailto:#2}{#2}}}\frontmatter@RRAPformat}
  {}{}
}%
\makeatother
\begin{document}

\preprint{AIP/123-QED}

\title[The Partially Saturated Cells Method for Complex Moving Geometries in Turbulent Flow Conditions]{The Partially Saturated Cells Method for Complex Moving Geometries in Turbulent Flow Conditions}

\author{Philipp Suffa}
\email{philipp.suffa@fau.de}
\affiliation{
Chair for System Simulation, Friedrich-Alexander-Universität Erlangen-Nürnberg, Erlangen, Germany
}
\author{Harald Köstler}
\affiliation{ 
Chair for System Simulation, Friedrich-Alexander-Universität Erlangen-Nürnberg, Erlangen, Germany
}
\author{Ulrich Rüde}
\affiliation{ 
Department for Applied Mathematics, VSB-Technical University of Ostrava, Ostrava, Czech Republic
}

\date{\today}

\begin{abstract}
We present four application cases to test and validate fluid-structure-interaction models based on the Lattice Boltzmann and the Partially Saturated Cells method (PSM) for turbulent flow conditions. 
For a flow around a stationary sphere we compare the PSM with direct mapping boundary conditions and achieve good agreement to reference data for Reynolds numbers up to one million. 
In particular, the model predicts the drag crisis for this case with good accuracy.
For a three dimensional airfoil simulation at a Reynolds number of $150000$, the PSM predicts the stall angle for a NACA0018 profile with excellent accuracy. 
For a geometry-resolved stirred tank simulation at a Reynolds number of $29000$ the PSM produces the time averaged velocity profiles around the impeller with excellent fidelity.
Finally, a wind turbine geometry is resolved with the PSM for a tip Reynolds number of $\sim 40000$, where thrust and mean velocity in the turbine wake shows excellent agreement with reference data from literature.  
\end{abstract}

\maketitle

\input{chapters/introduction}

\input{chapters/method}

\input{chapters/flowAroundSphere}

\input{chapters/airfoil}

\input{chapters/stirredTank}

\input{chapters/turbine}

\input{chapters/conclusion}

\section*{Acknowledgements}
This project has received funding from the European High Performance Computing Joint Undertaking under grant agreement n\textdegree\,101144014.
The research of the 3rd author was partially funded by the European Union under the REFRESH -- Research Excellence For Region Sustainability and High-tech Industries project number CZ.10.03.01/00/22\_003/0000048 via the Operational Programme Just Transition.
The authors gratefully acknowledge the Gauss Centre for Supercomputing e.V. (\url{www.gauss-centre.eu}) for funding this project by providing computing time through the John von Neumann Institute for Computing (NIC) on the GCS Supercomputer JUWELS at Jülich Supercomputing Centre (JSC). 
We acknowledge the EuroHPC Joint Undertaking for awarding this project access to the EuroHPC supercomputer LUMI, hosted by CSC (Finland) and the LUMI consortium through a EuroHPC Regular Access call. 
The authors gratefully acknowledge the scientific support and HPC resources provided by the Erlangen National High Performance Computing Center (NHR@FAU) of the Friedrich-Alexander-Universität Erlangen-Nürnberg (FAU).

\section*{Author contributions}
\noindent \textbf{P. Suffa:} Conceptualization, Methodology, Software, Validation, Formal analysis, Investigation, Data Curation, Writing - Original Draft, Visualization, Project administration \\
\textbf{H. Köstler:} Writing - Review \& Editing, Supervision, Funding acquisition \\
\textbf{U. Rüde:} Writing - Review \& Editing, Supervision, Funding acquisition \\

\section*{Conflict of Interest Statement}
The authors have no conflicts to disclose.

\section*{Generative AI disclosure}
During the preparation of this manuscript, the authors used Claude (Anthropic) to draft and improve the readability and language of the text, while all structure and content were provided by the authors. 
No technical claims, figures or interpretations are AI-generated. 
The authors reviewed and edited all content and take full responsibility for it.

\section*{Data Availability Statement}
The data that support the findings of this study are openly available on Zenodo: \url{https://doi.org/10.5281/zenodo.22897962}

\section*{References}
\bibliography{aipsamp}

\end{document}

%% file: chapters/introduction.tex
\section{Introduction}

\begin{figure}
    \centering
    \includegraphics[width=\linewidth, trim={4cm 0cm 6cm 0cm},clip]{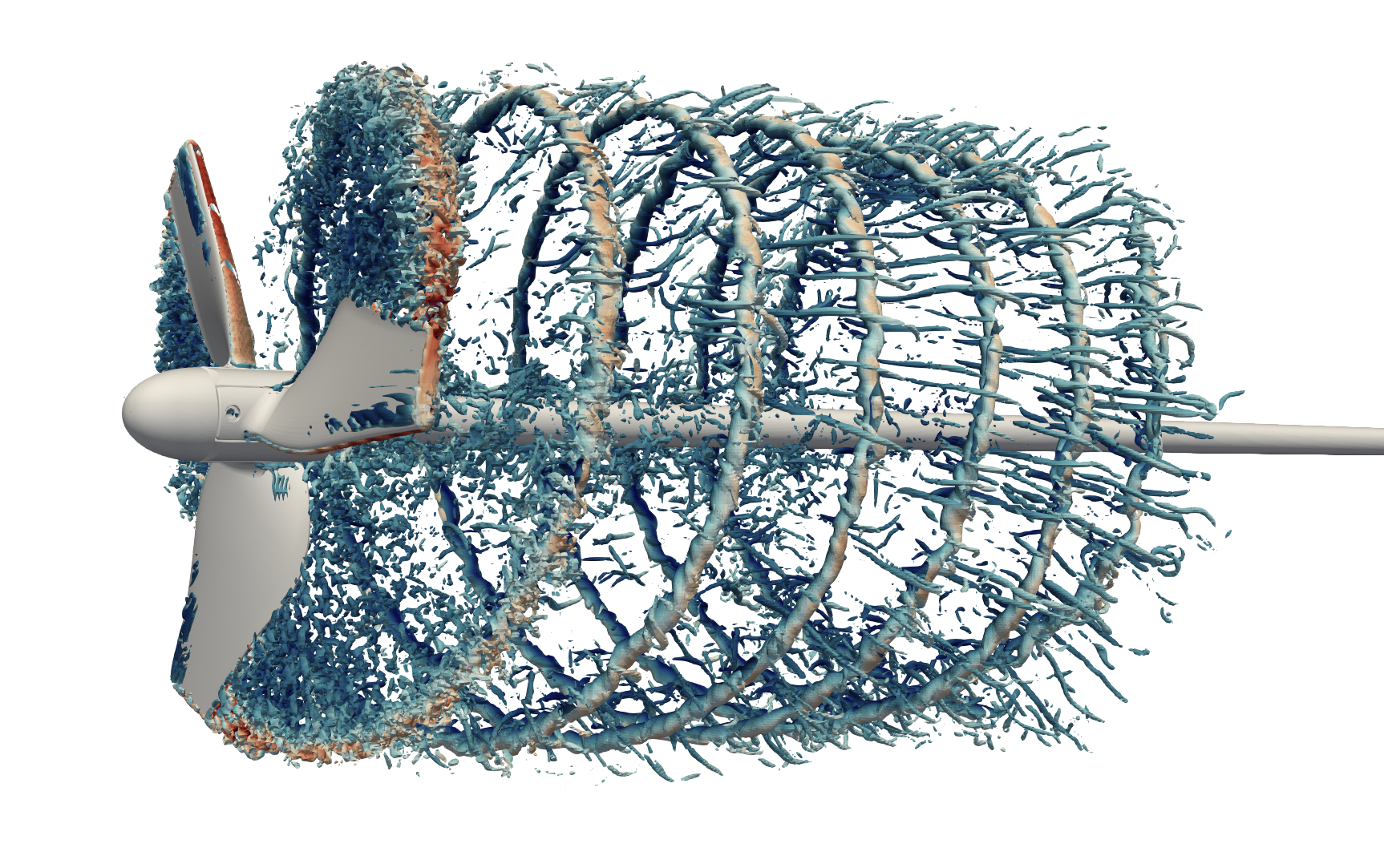}
    \caption{The turbulent flow around a blade-resolved rotor, visualized by the Q-criterion.}
    \label{fig:turbine_pic}
\end{figure}

Wind turbines, stirred reactors, and rotor systems are typical examples of industrial applications in which an object having complex geometry moves through a surrounding fluid.
The efficient and accurate simulation of such configurations is therefore of great practical importance.

The lattice Boltzmann method (LBM) has established itself as an accurate approach for the simulation of fluid flow.
Since it is an explicit time-stepping scheme operating on a Cartesian lattice, it exposes a high degree of node-level and instruction-level parallelism.
For this reason, the LBM is exceptionally well suited for GPUs and for large-scale simulations on typical current supercomputing architectures (\cite{watanabePerformanceEvaluationLattice2022}, \cite{liuAcceleratingLargeScaleCFD2023}, \cite{spinelliHPCPerformanceStudy2023}, \cite{holzerHighlyEfficientLattice2021}).

If the geometry moves in the flow field, the flow solver must additionally provide a fluid-structure interaction (FSI) mechanism that couples the object motion to the fluid. 
Several strategies have been established for this purpose in the LBM context.

A well-established approach is the momentum exchange method (MEM), which goes back to the work of \cite{ladd1994numerical}.
In the MEM, the object is mapped explicitly onto the lattice, and its influence on the fluid is imposed through boundary conditions at the solid surface, ranging from simple bounce-back schemes to higher-order interpolated variants (\cite{Ginzburg2008427}).
In the reverse direction, the hydrodynamic force and torque acting on the object are obtained from the momentum that the particle distribution functions (PDFs) transfer across the fluid-solid interface.
The MEM has been applied successfully to a wide range of fully resolved particulate flows (\cite{gotz2010direct}, \cite{rettingerComparativeStudyFluidparticle2017a}, \cite{werner2021coupling}).
For moving objects, however, the explicit mapping implies that lattice cells previously covered by the solid become part of the fluid domain, and their missing PDFs must be reconstructed before the next update.
This so-called {\em fresh-node} problem (\cite{lbm_book}) requires dedicated refilling schemes (\cite{schwarzmeier2022comparison}), whose design is delicate: an inappropriate treatment can degrade both accuracy and stability (\cite{rettingerEfficientFourwayCoupled2022}).

In the immersed boundary method (\cite{fengImmersedBoundarylatticeBoltzmann2004}), the effect of the solid surface is imposed on the fluid through body forces, which makes the approach applicable to arbitrarily moving objects and particles.
Sliding-mesh techniques (\cite{zhang2011}) instead couple a rotating grid region to a stationary one; strictly speaking, they are not an FSI method but a way of representing a rotating geometry on the lattice, while the interaction with the solid surface is still realized by classical boundary conditions.
In wind-energy applications, actuator methods are the most common choice: they do not resolve the blade geometry at all, but model its effect on the flow by equivalent force distributions (\cite{asmuthActuatorLineSimulations2020,schottenhammlWalberlawindLatticeBoltzmannbased2024}).

In this study, we employ the Partially Saturated Cells Method (PSM), introduced by \cite{nobleLatticeBoltzmannMethodPartially1998a}, to represent fully resolved complex moving geometries in LBM solvers.
To date, the method has mainly been used for fluid-particle systems with geometrically resolved particles (\cite{majumderReexaminingPartiallySaturatedcells2022a}, \cite{rettingerComparativeStudyFluidparticle2017a}), and it has been demonstrated that it is not restricted to spherical particles but can also represent arbitrarily shaped objects (\cite{trunkSimulationArbitrarilyShaped2018}).
A distinctive feature of the PSM is that solid objects do not require any explicit boundary handling: the no-slip condition at the object surface is incorporated implicitly into the LBM update itself.
The method therefore preserves the structure, and with it the inherent parallelism, of the plain LBM, which makes it an attractive choice for GPU-based simulations with explicit fluid-solid coupling (\cite{kemmlerEfficiencyScalabilityFullyresolved2025}, \cite{benseghierParallelGPUbasedComputational2020}).
As an additional benefit, no special treatment of fresh fluid nodes that are uncovered by the moving geometry is needed.

The accuracy of the PSM has been examined in a number of studies.
Second-order convergence was already reported for a moving cylinder in the original work (\cite{nobleLatticeBoltzmannMethodPartially1998a}) and was later confirmed for the force on a periodic array of spheres (\cite{rettingerComparativeStudyFluidparticle2017a}).
The authors of \cite{majumderReexaminingPartiallySaturatedcells2022a} revisited the method for stationary, oscillating, and rotating cylinders: for the stationary cylinder in unsteady flow at Reynolds numbers up to 200, the computed drag and lift coefficients agreed excellently with literature data, while for moving objects, second-order accuracy was retained for the pressure but the velocity degraded to first order.
Good agreement of drag and lift coefficients with reference results was likewise reported for a stationary cylinder in \cite{trunkSimulationArbitrarilyShaped2018}.
Finally, the PSM was found to reproduce vortex-induced vibrations of an eccentrically positioned cylinder in a transient Couette flow correctly (\cite{haussmannGalileanInvarianceStudy2020}).

This work is a continuation of \cite{suffaLargescaleSimulationsFully2025}, where we introduced the methodology of the PSM and its integration into the framework \walberla (\cite{waLBerla}).
There, we further showed a physical validation for laminar flow problems, such as a settling sphere and a rotating cylinder test in the low Reynolds regime.
Lastly, excellent node-level and scaling performance of the PSM on CPU and GPU was presented for an industrial application, namely a counter-rotating open rotor.

The present study builds on this previous work and demonstrates the capabilities of the PSM for simulations of complex moving geometries in turbulent flow conditions.

The main contributions of this study are the
\begin{enumerate}
    \item prediction of the drag crisis for a flow around a stationary sphere resolved by the PSM, and a comparison to direct mapping boundary conditions;
    \item stall prediction for the NACA0018 airfoil with a variety of angles of attack, and a comparison of boundary conditions;
    \item simulation and validation of a stirred mixer with a fully-resolved moving geometry realized with the PSM;
    \item simulation and validation of a fully-resolved rotor geometry.
\end{enumerate}

%% file: chapters/method.tex
\section{Method}
\subsection{The Partially Saturated Cells Method}
In the following, a brief recap of the PSM is given. For a more in-depth introduction, see the original work from \cite{nobleLatticeBoltzmannMethodPartially1998a}.

The PSM extends the classical LBM by introducing a coupling mechanism that
transfers momentum between the fluid and the solid phase.
In the LBM, the simulation domain is discretized on a Cartesian lattice, where
every cell stores a set of particle distribution functions (PDFs) that evolve according to a time-stepping discretization of the Boltzmann equation.
One time step of the method can be written as
\begin{equation}
    f_i(\textbf{x} + \Delta t \textbf{c}_i, t+ \Delta t) = f_i(\textbf{x},t) + \Omega_i^F,
\end{equation}
where $f_i(\textbf{x},t)$ denotes the PDF stored in cell $\textbf{x}$ for
lattice direction $i$, and $\textbf{c}_i$ is the corresponding discrete
velocity. The operator $\Omega_i^F$ models the collision of particle
distributions by relaxing the PDFs towards a local equilibrium.
The simplest choice for this operator is the Single Relaxation Time scheme
(SRT, \cite{qianLatticeBGKModels1992}), given by
\begin{equation}
    \Omega_i^F = - \frac{1}{\tau}(f_i(\textbf{x},t) - f^{(\text{eq})}_i(\textbf{x},t)),
\end{equation}
where $\tau$ denotes the relaxation time.
Throughout this work, however, we primarily employ the Cumulant collision
operator (\cite{GEIER2015507}), an advanced scheme that offers improved accuracy
and stability, in particular for turbulent flow regimes.

To obtain the PSM, the lattice Boltzmann equation is augmented with an
additional solid collision operator $\Omega_i^S$, whose contribution is
weighted by a solid volume fraction field $B(\textbf{x}, t)$:
\begin{multline}
        f_i(\textbf{x} + \Delta t \textbf{c}_i, t + \Delta t) = \\ f_i(\textbf{x},t) + (1 - B(\textbf{x}, t)) \cdot \Omega_i^F + B(\textbf{x}, t) \cdot \Omega_i^S.
        \label{equ:PSM_rule}
\end{multline}
The scalar field $B(\textbf{x}, t)$ encodes to what extent cell $\textbf{x}$
is occupied by a solid body and thereby controls the relative weight of the
fluid and solid collision terms in \autoref{equ:PSM_rule}.

\begin{figure}
\centering
\includegraphics[width=0.85\linewidth, trim={10cm 9.5cm 13cm 2.5cm},clip]{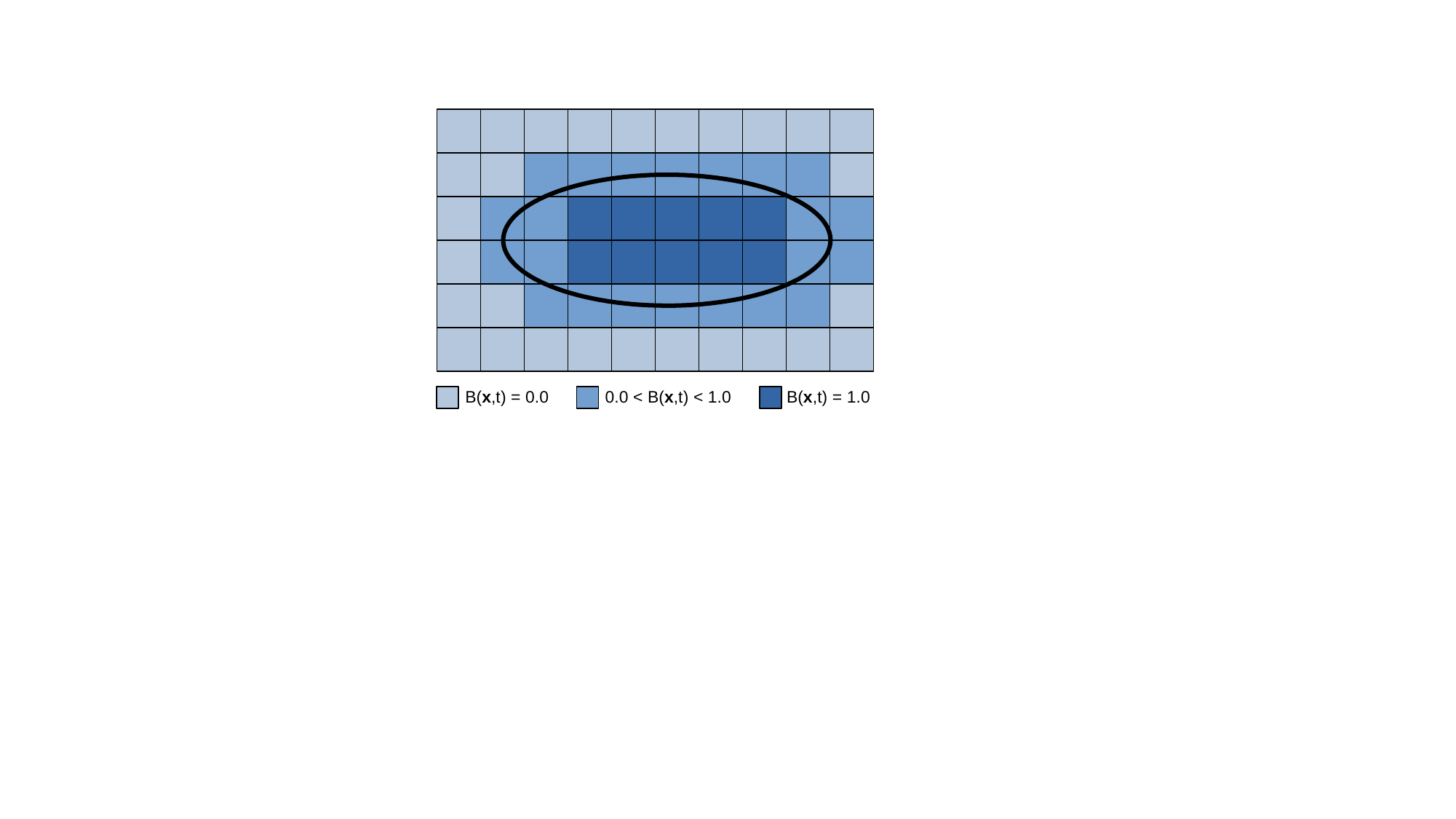}
\caption{\label{fig:fractionField} The fraction field $B(\textbf{x},t)$ for the Partially Saturated Cells method representing an example geometry of an ellipse. From \cite{suffaLargescaleSimulationsFully2025}}
\end{figure}

\autoref{fig:fractionField} illustrates this field for an elliptical object:
cells located entirely outside the ellipse carry a value of
$B(\textbf{x}, t) = 0$, cells fully covered by the object carry
$B(\textbf{x}, t) = 1$, and cells intersected by the boundary take
intermediate values proportional to the fraction of the cell volume that
overlaps with the body.
An efficient evaluation of $B(\textbf{x}, t)$ is a non-trivial aspect of the
PSM (\cite{lbm_book}); this holds in particular for moving objects, where the
field has to be recomputed repeatedly over the course of the simulation.
An efficient strategy for initializing and updating $B(\textbf{x}, t)$ for
moving geometries is described in detail in
\cite{suffaLargescaleSimulationsFully2025}.

The solid collision operator $\Omega_i^S$ in \autoref{equ:PSM_rule} functions as a no-slip boundary condition and yields
\begin{equation}
        \Omega_i^S = \left[f_i^{(\text{eq})}(\rho, \textbf{u}_s) - f_i(\textbf{x},t) \right] + \left( 1 - \frac{1}{\tau} \right) \left[ f_i(\textbf{x},t) - f_i^{(\text{eq})}(\rho,\textbf{u}_s) \right].
    \label{equ:solid_collision2}
\end{equation}

As can be seen from \autoref{equ:PSM_rule}, the two limiting cases of the
solid volume fraction recover distinct behaviors: for $B(\textbf{x}, t) = 0$
the update reduces to a standard LBM collision, whereas for
$B(\textbf{x}, t) = 1$ the equilibrium distributions are bounced back.
In cells that are only partially covered by the solid, i.e.,
$0 < B(\textbf{x}, t) < 1$, both mechanisms act simultaneously in a blended
fashion.
In this way, the effect of the solid body on the surrounding flow is
accounted for.
For the reverse coupling, i.e., the feedback of the fluid onto the solid, the
hydrodynamic force acting on the object is accumulated as
\begin{equation}
    \textbf{F}(t) = \frac{\Delta x^3}{\Delta t} \sum_{\textbf{x}_s} B(\textbf{x}_s, t) \sum_i \Omega_i^S \textbf{c}_i ,
    \label{equ:forcing}
\end{equation}
together with the corresponding torque
\begin{equation}
    \textbf{T}(t) = \frac{\Delta x^3}{\Delta t} \sum_{\textbf{x}_s}  B(\textbf{x}_s, t) (\textbf{x}_s - \textbf{R}) \times \sum_i \Omega_i^S \textbf{c}_{i},
    \label{equ:torque}
\end{equation}
where the summation runs over all cells $\textbf{x}_s$ that intersect the
object, and $\textbf{R}$ denotes its center of mass (\cite{lbm_book}).

\subsection{Direct Mapping Boundary Conditions for Resting Objects}
\label{sec:direct_mapping}
In this section, we briefly introduce the no-slip bounce-back and the Quadratic-Bounce-Back (QBB) boundary condition, which are later used for comparison with the PSM.

In the LBM, the no-slip boundary condition is typically enforced through a bounce-back scheme, which reverses the momentum of particle distribution functions at solid surfaces to impose zero tangential velocity.
For a boundary node $x_b$, and a lattice direction $i$ with opposite direction $\bar{i}$, the PDF $f_i$ in the next time step $t + \Delta t$ is set to the post-collision population $f^*$ as
\begin{equation}
    f_{\bar{i}}(\textbf{x}_b, t+\Delta t) = f^*_{i}(\textbf{x}_b,t).
    \label{equ:noslip}
\end{equation}
In the widely used half-way bounce-back formulation, the wall is located midway between fluid and solid nodes, leading to a staircase approximation of the boundary and thus to a boundary condition that is only first-order accurate for complex shapes.
For further details about the no-slip boundary condition, see \cite{lbm_book}.

The Quadratic-Bounce-Back boundary condition is a single-node interpolation scheme for the simple bounce-back boundary condition.
It is similar to the velocity boundary condition concept proposed by \cite{Bouzidi2001}, and therefore second-order accurate.
The benefits of the QBB condition compared to \cite{Bouzidi2001} are that it can be implemented without branching, does not need access to neighboring cells, and avoids storing both pre- and post-collision PDF states.
The simplified version for stationary boundaries reads
\begin{equation}
\begin{split}
    f_{\bar{i}}(\textbf{x}_b, t+\Delta t) &= \frac{1}{q+1} ((1-q)f_{i}(\textbf{x}_b,t) + qf^*_{i}(x_b,t)) \\ 
    &+ \frac{q}{q+1} f^*_ {\bar{i}}(\textbf{x}_b,t),
\end{split}
\label{equ:qbb}
\end{equation}
with $q$ as the distance from the cell center of $x_b$ to the actual geometrical boundary.
In \autoref{equ:qbb}, the pre-collision PDF $f_{i}(x_b,t)$ is needed. 
As we only store post-collision PDFs, the pre-collision state must be recovered from the post-collision state based on a single relaxation time collision operator
\begin{equation}
\begin{split}    
f_{i}(\textbf{x}_b,t) &\approx \frac{f^*_i(\textbf{x}_b,t) - f^*_{\bar{i}}(\textbf{x}_b,t)}{2} \\ &+ \frac{f^*_i(\textbf{x}_b,t) + f^*_{\bar{i}}(\textbf{x}_b,t) - \omega_1 (f^{\text{eq}}_i(\textbf{x}_b,t) + f^{\text{eq}}_{\bar{i}}(\textbf{x}_b,t)) }{2 - 2 \omega_1}.
\end{split}
\end{equation}
Using an SRT operator for this recovery also works when a different collision operator is used in the LBM collision step.
For more details about the QBB boundary condition, see \cite{GEIER2015507}.

Both boundary conditions can be extended to support moving geometries; however, this leads to the so-called {\em fresh-node problem} in both cases, see, e.g., \cite{schwarzmeier2022comparison}.
Because both boundary conditions are based on a direct-mapping approach,
cells covered by the solid geometry hold no valid fluid data.
Thus, if the object moves and a former boundary cell must be converted into a fluid cell, its PDF information must be restored by a dedicated mechanism.
Such mechanisms are complex, may induce instabilities, and can impact the performance of the application, especially on graphics processing units.
Avoiding the fresh node problem is one of our main reasons to study the PSM in detail.

In the following applications, forces are calculated on the geometry objects. 
For the no-slip and QBB boundaries, the forces are calculated by a momentum-exchange approach at the surface of the object.
The total force on an object boundary can be calculated by
\begin{equation}
    \textbf{F}(t) = \frac{\Delta x^3}{\Delta t} \sum_{x_b} \sum_i (f_i^{\text{in}} + f_{\bar{i}}^{\text{out}}) \textbf{c}_i, 
\label{equ:qbb_force}
\end{equation}
with $f_i^{\text{in}}$ as incoming PDF and $ f_{\bar{i}}^{\text{out}}$ as bounced-back PDF from \autoref{equ:noslip} or \autoref{equ:qbb} (\cite{lbm_book}).

%% file: chapters/flowAroundSphere.tex
\section{Flow around a sphere: Drag crisis}
The overall goal of this work is to apply the PSM to moving geometries in highly turbulent flows.
As a first step, however, we consider a stationary object, which allows us to assess the PSM as a representation of solid boundaries in turbulent flow independently of the additional challenges introduced by object motion.
So far, the PSM has mainly been employed for particulate and porous-media flows, and its validation has been limited to laminar or moderately unsteady conditions (\cite{suffaLargescaleSimulationsFully2025}); to the best of our knowledge, its accuracy as a boundary representation in highly turbulent flows has not yet been investigated.
To test the method under particularly demanding conditions, we therefore simulate the flow around a sphere across the drag crisis and compare the PSM to direct mapping boundary conditions.

The drag crisis is a Reynolds-number–dependent transition in the flow around a sphere that produces a sudden, large reduction in the drag coefficient and is driven by boundary-layer transition; it therefore has major consequences for force prediction, wake dynamics, and both experimental and numerical modeling. 
For smooth spheres the critical Reynolds number lies near $\sim 3\times 10^5$, while surface roughness and upstream turbulence can shift the transition substantially, because an earlier laminar-to-turbulent boundary-layer transition delays separation, narrows the wake, and reduces pressure drag. 
The phenomenon was noted early in the literature (\cite{eiffel1914nouvelles,GDZPPN002503859}) and has been characterized in numerous experimental studies (\cite{Roshko_1961,achenbachExperimentsFlowSpheres1972,cadot2015}), which demonstrate the strong sensitivity of the critical Reynolds number and the magnitude of drag reduction to surface condition and inflow disturbances. 
The drag crisis matters for applications for three main reasons: it introduces a non-monotonic, strongly nonlinear dependence of aerodynamic forces on the Reynolds number, so neglecting the crisis can produce force errors of tens of percent with direct impact on trajectory, stability, and energy estimates; 
by altering wake structure and vorticity transport it modifies mixing, acoustic signatures, and unsteady loading, affecting environmental dispersion, sediment transport, and noise generation; 
and it poses a stringent validation target for numerical methods, since capturing the transition and its sensitivity requires either scale-resolving simulations or turbulence models and meshing strategies that have been carefully validated against experiments (\cite{kawamuraComputationHighReynoldsb,mutoNegativeMagnusLift2012,constantinescuNumericalInvestigationsFlow2004,lehmkuhlUnsteadyForcesCircular2014,chopraIntermittentNatureLaminar2017}). 
Recent lattice-Boltzmann studies further illustrate both the promise and the challenges of alternative numerical approaches for resolving the crisis (\cite{namburiCrystallographicLatticeBoltzmann2016,holzerCodeGenerationLattice2025,geierParametrizationCumulantLattice2017a}).

\begin{figure*}
\centering
\includegraphics[width=0.8\linewidth, trim={2cm 2cm 3cm 1cm},clip]{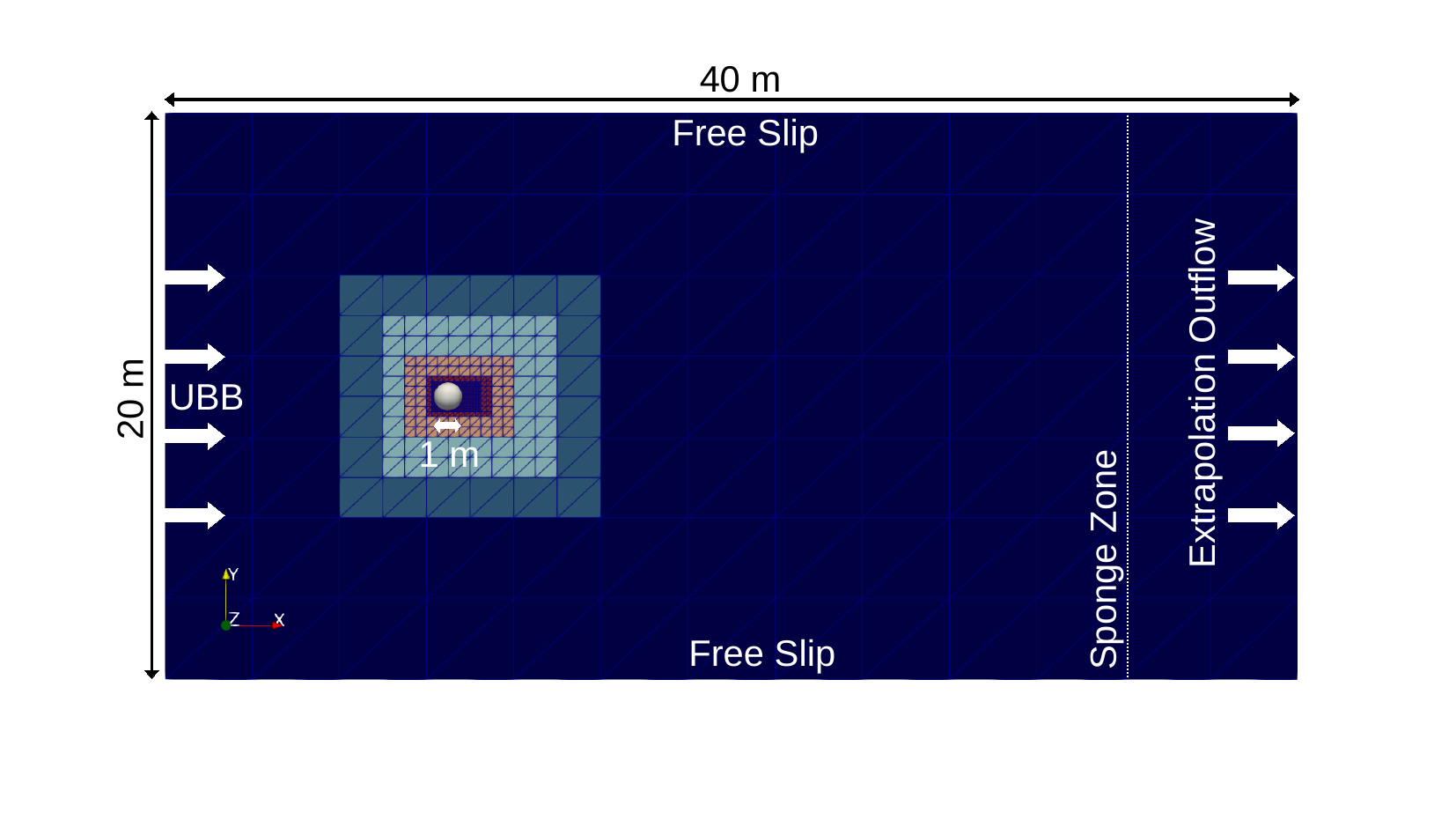}
\caption{Domain decomposition and boundary setup for the flow around sphere simulations for Reynolds numbers from $10^5$ to $10^6$.}
\label{fig:domain}
\end{figure*}

\begin{figure*}
     \centering
         \includegraphics[width=\textwidth, trim={5cm 33.5cm 5cm 0cm},clip]{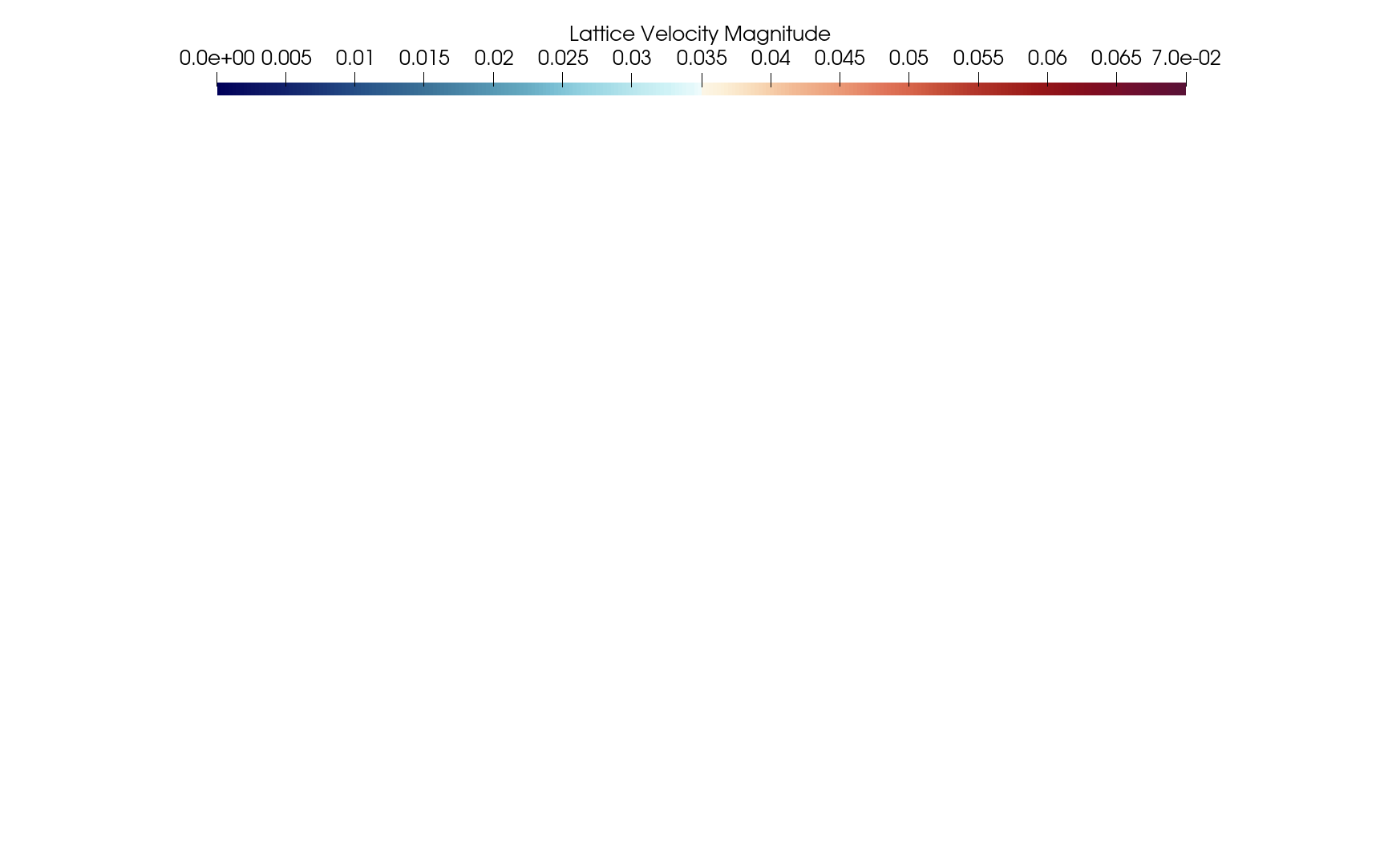}
     \begin{subfigure}[b]{0.45\textwidth}
         \centering
         \includegraphics[width=\textwidth]{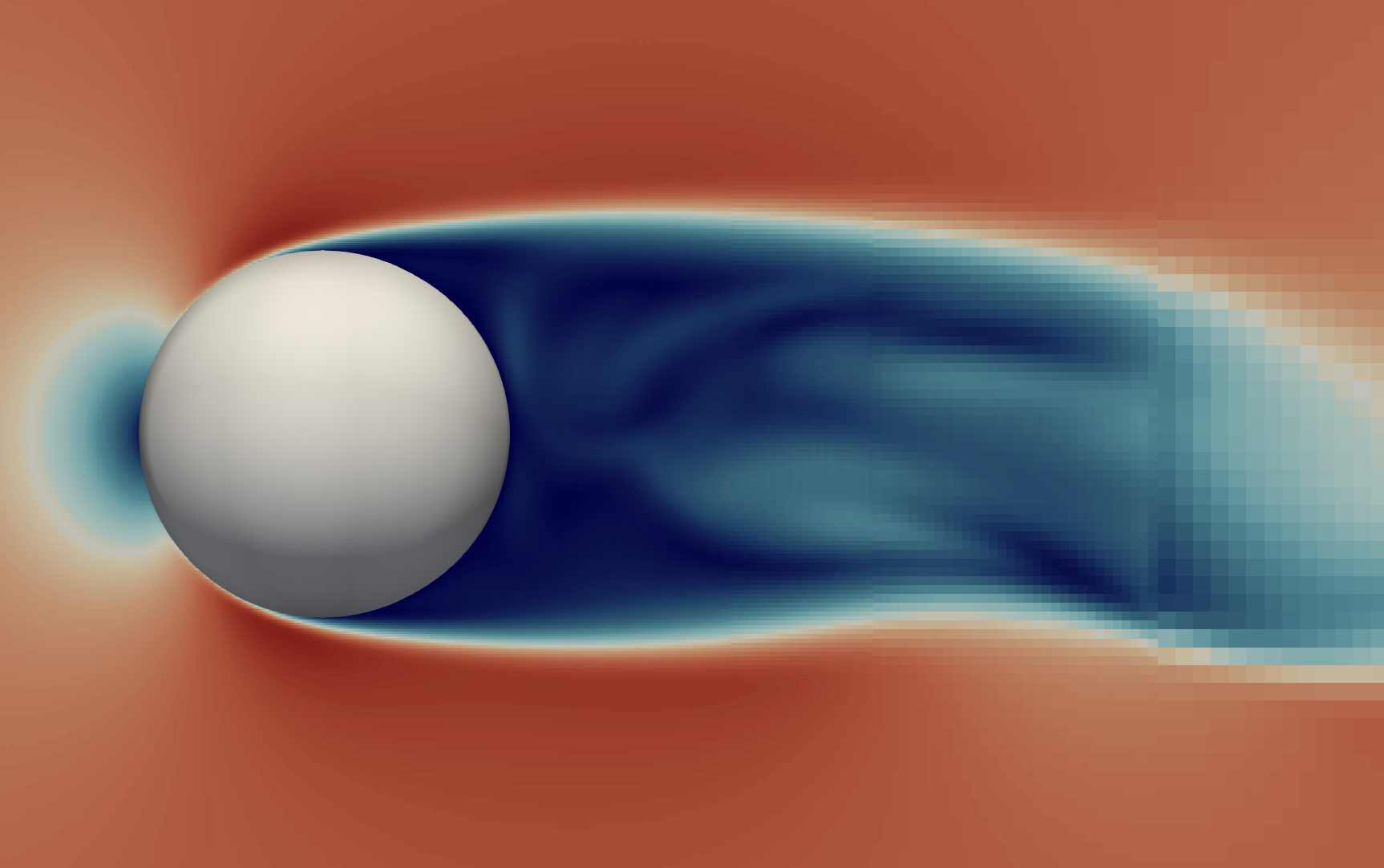}
         \caption{Velocity Magnitude for Reynolds Number 1,000.}
         \label{fig:vel1k}
     \end{subfigure}
     \begin{subfigure}[b]{0.45\textwidth}
         \centering
         \includegraphics[width=\textwidth]{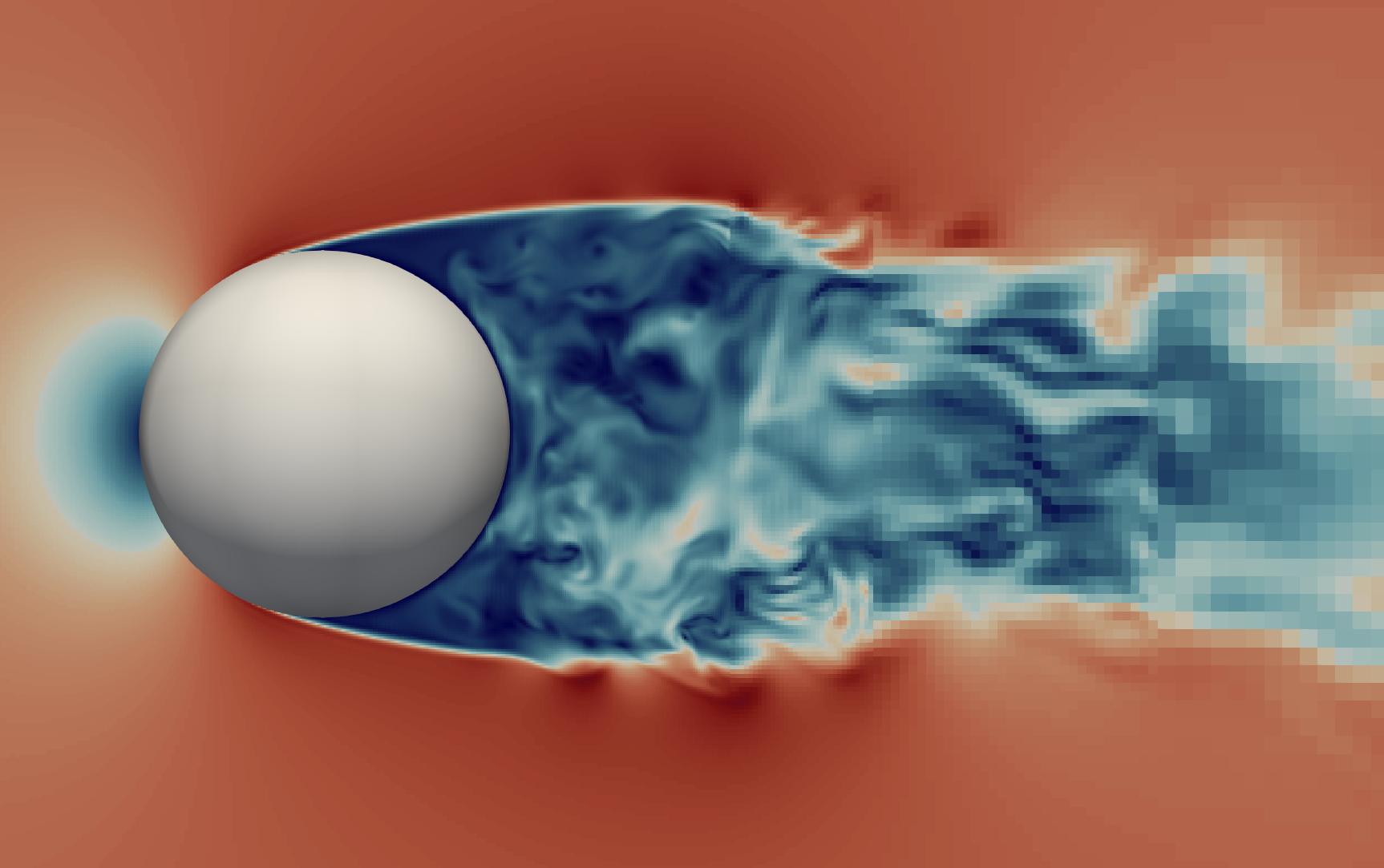}
         \caption{Velocity Magnitude for Reynolds Number 10,000.}
         \label{fig:vel10k}
     \end{subfigure}
     \begin{subfigure}[b]{0.45\textwidth}
         \centering
         \includegraphics[width=\textwidth]{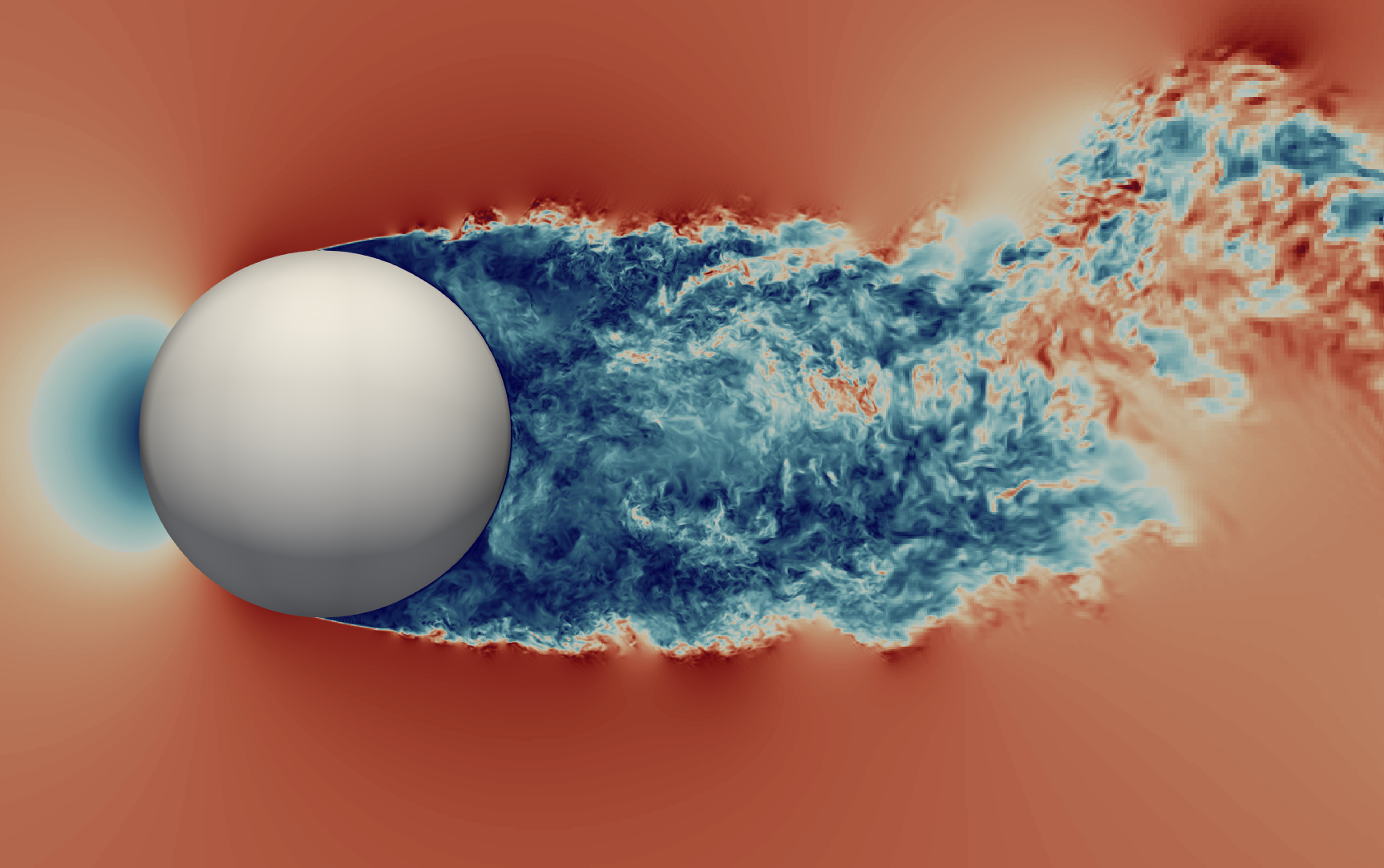}
         \caption{Velocity Magnitude for Reynolds Number 100,000.}
         \label{fig:vel100k}
     \end{subfigure}
     \begin{subfigure}[b]{0.45\textwidth}
         \centering
         \includegraphics[width=\textwidth]{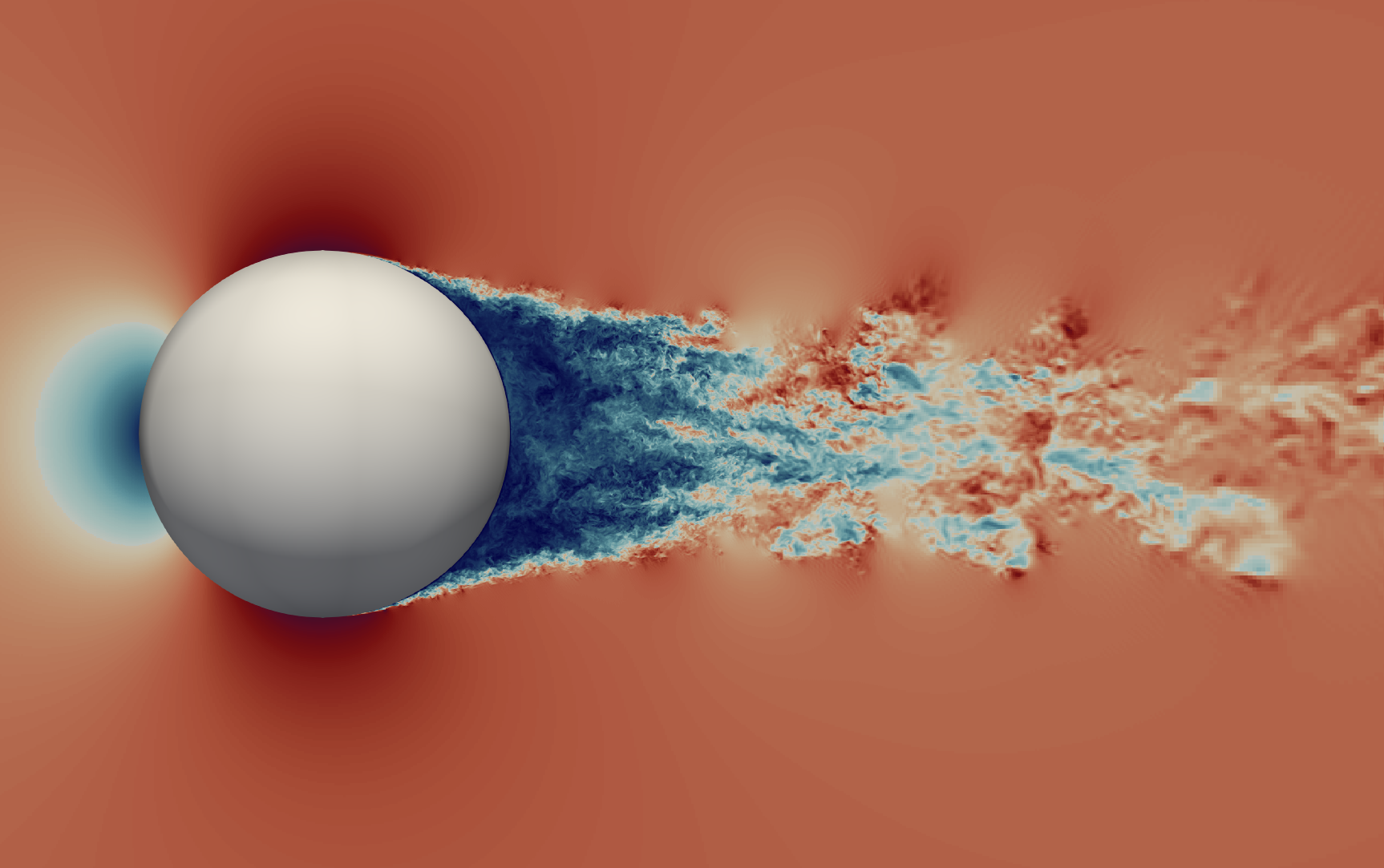}
         \caption{Velocity Magnitude for Reynolds Number 500,000.}
         \label{fig:vel500k}
     \end{subfigure}
        \caption{Snapshot of the instantaneous velocity field around the sphere for different Reynolds numbers. The boundary condition of the sphere is handled by the PSM.}
        \label{fig:velFields}
\end{figure*}

\subsection{Simulation setup}

We perform an extended validation of the PSM for the flow around a sphere for Reynolds numbers ranging from $10^{-4}$ to $10^6$.
The Reynolds number is defined as 
\begin{equation}
Re = \frac{u_{0} \cdot L_{\text{lb}}}{\nu_{\text{lb}}}, 
\end{equation}
with $u_{0}$ as inflow velocity in lattice units, $L_{\text{lb}}$ as sphere diameter in lattice cells and $\nu_{\text{lb}}$ as kinematic viscosity in lattice units.
The output of our simulations is an averaged drag coefficient 
\begin{equation}
    C_D = \frac{8F_x}{\rho u_0^2 \pi d^2}, 
\label{equ:drag}
\end{equation}
with $F_x$ as integrated force component in axial direction, $\rho$ as fluid density and $d = L_{\text{lb}}$ as sphere diameter.
Integrating the force components over the whole geometry with \autoref{equ:forcing} can be expensive, especially because atomic operations are involved on GPUs.
Therefore, we calculate forces only every ten coarse time steps for all simulations in this work.

For a large range of Reynolds numbers, different domain setups and simulation parameters are needed to ensure a good compromise between numerical accuracy and computational cost. 
In \autoref{tab:domain}, the different domain and simulation parameters are presented.
The sphere has a diameter of one meter in physical units.
For the flow around a sphere in laminar flow conditions, i.e., for Reynolds numbers $<10^2$, a sphere diameter resolution of 10 cells and a domain size of $11\text{ m}\times11\text{ m}\times11\text{ m}$ are sufficient.
Further, we use the SRT collision operator and a D3Q19 stencil in this regime.

Other parameters that vary with the test case are the inflow velocity $u_0$ and the relaxation rate $\omega$.
Over such a wide range of Reynolds numbers, it is crucial to adapt $u_0$ and $\omega$ to the Reynolds number.
If $u_0$ is too small, the simulation converges very slowly to steady state.
If $u_0$ is too large, the simulation becomes inaccurate or even unstable.

We differentiate between simulations with Reynolds numbers $<1$, and simulations with $Re \geq 1$.
For $Re < 1$, $\omega = 0.55$ and $u_0$ is calculated from 
\begin{equation}
    u_0 = \frac{\nu_{\text{lb}} Re}{L_{\text{lb}}}, 
\end{equation}
with the kinematic viscosity in lattice units $\nu_{\text{lb}}$ calculated from the relaxation rate $\omega$ as 
\begin{equation}
    \nu_{\text{lb}} = \frac{1}{3} (\frac{1}{\omega} - 0.5).
\end{equation}

For higher Reynolds numbers, this approach would yield an inflow velocity larger than 0.1, which leads to poor accuracy and instabilities (\cite{lbm_book}).
Therefore, we set $u_0= 0.05$ for $Re \geq 1$ and calculate the relaxation rate with
\begin{equation}
\omega = 1 / (0.5 + 3 \nu_{\text{lb}}) = 1 / (0.5 + 3 \frac{u_{0} \cdot L_{\text{lb}}}{Re}).
\end{equation}

For turbulent flow conditions, we use the Cumulant collision operator and a D3Q27 stencil for increased accuracy and stability in these conditions (\cite{GEIER2015507}). We also increase the domain size to $40\text{ m} \times 20\text{ m} \times 20\text{ m}$ to decrease the influence of the boundary conditions on the forces on the sphere.

For all simulations, a velocity bounce back boundary is set as inflow in western direction, a simple extrapolation outflow boundary (see \cite{GEIER2015507}) is set in eastern direction, and all other boundaries are handled by a free-slip boundary condition. 

In \autoref{fig:domain}, the domain for the most demanding runs with Reynolds numbers from $10^5$ to $10^6$ is displayed. 
For simulations in this Reynolds number regime, a sponge zone is set next to the outflow boundary, to improve the stability of the simulation. 
In the sponge zone, the relaxation rate decreases from its original level at the start of the sponge zone to a target relaxation rate $\omega_t = 1.9$ at the outflow boundary.
Further, the refinement levels of the simulation domain are displayed.
A maximum of five refinement levels with a mesh resolution of 0.001 m is reached in the region of interest, i.e., the area around the sphere.
The simulation domain reaches a total of $2.62\cdot 10^9$ LBM cells, and runs for $2.1 \cdot 10^6$ time steps on the finest mesh level.
This time is needed for the flow to cross the domain in flow direction approximately twice, and to fully develop the flow and converge to a stable drag coefficient.
The simulations with the largest domain setup run for up to 96 hours on 32 NVIDIA H100 GPUs on the Helma HPC system\footnote{\url{https://doc.nhr.fau.de/clusters/helma/}}, so one single simulation consumes more than 3000 GPU hours.

\begin{table*}
\caption{Parameters for flow around sphere simulations for a large range of Reynolds numbers.}
\label{tab:domain}
\begin{center}
\begin{tabular}{ c||c|c|c|c } 
 Re                         & $10^{-4}$ - $10^1$    & $10^2$        & $10^3$ - $10^4$  & $10^5$ - $10^6$           \\ \hline \hline
 Sphere Dia [\#cells]       & 10                    & 80            & 250              & 1000               \\ \hline
 Resolution [m]             & 0.1                   & 0.0125        & 0.004            & 0.001         \\ \hline
 Domain [m]                 & \multicolumn{2}{c|}{11x11x11}                            & \multicolumn{2}{c}{40x20x20}          \\ \hline
 Refinement levels          & 0                     & 2             & 4                & 5             \\ \hline
 Problem size [\#cells]     & $2 \cdot 10^6$        & $2\cdot 10^7$ & $1.6\cdot 10^8$  & $2.62\cdot 10^9$  \\ \hline
 Collision operator         & SRT                   & \multicolumn{3}{c}{Cumulant}     \\ \hline
 Fine \# time steps         & $10^4$                &$8.9\cdot 10^4$& $7.5 \cdot 10^5$ & $ \sim 2.1 \cdot 10^6$    \\ \hline
 GPU hours (NVIDIA H100) [h]              & <1                    & 2             & 96               & 3072
\end{tabular}
\end{center}
\end{table*}

\begin{figure*}
\centering
\includegraphics[width=\linewidth]{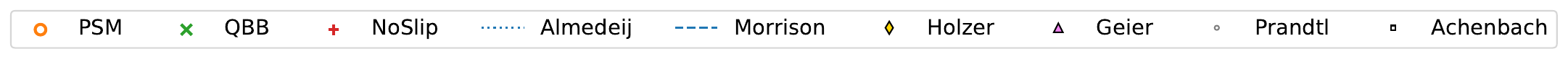}
\begin{subfigure}[b]{0.45\textwidth}
\includegraphics[width=\linewidth]{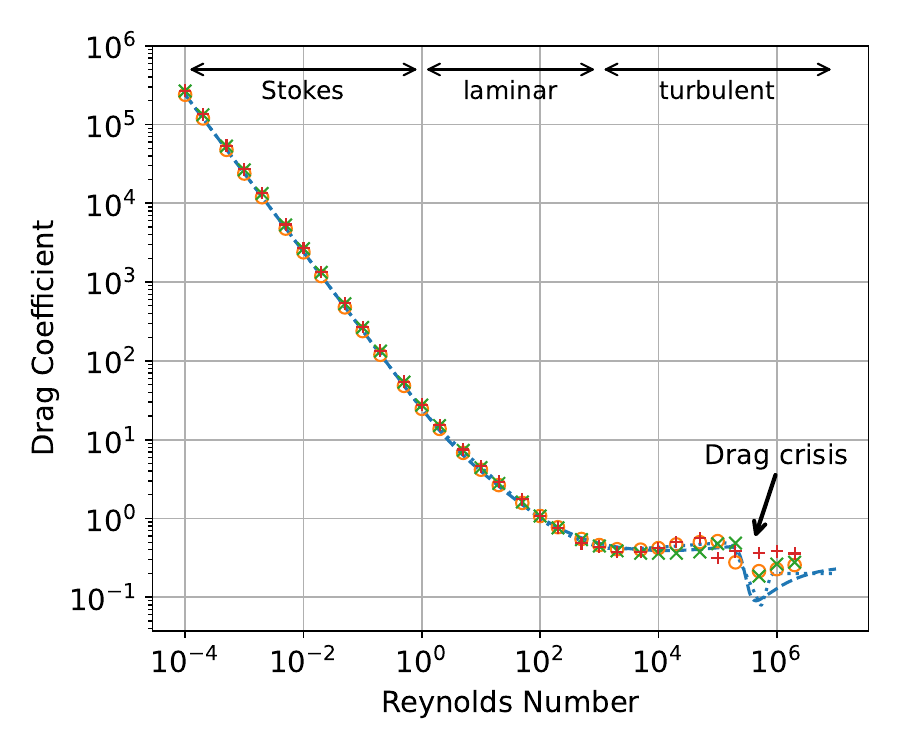}
\subcaption{Full Reynolds number range.}
\label{fig:sphere_low_re}
\end{subfigure}
\begin{subfigure}[b]{0.45\textwidth}
\includegraphics[width=\linewidth]{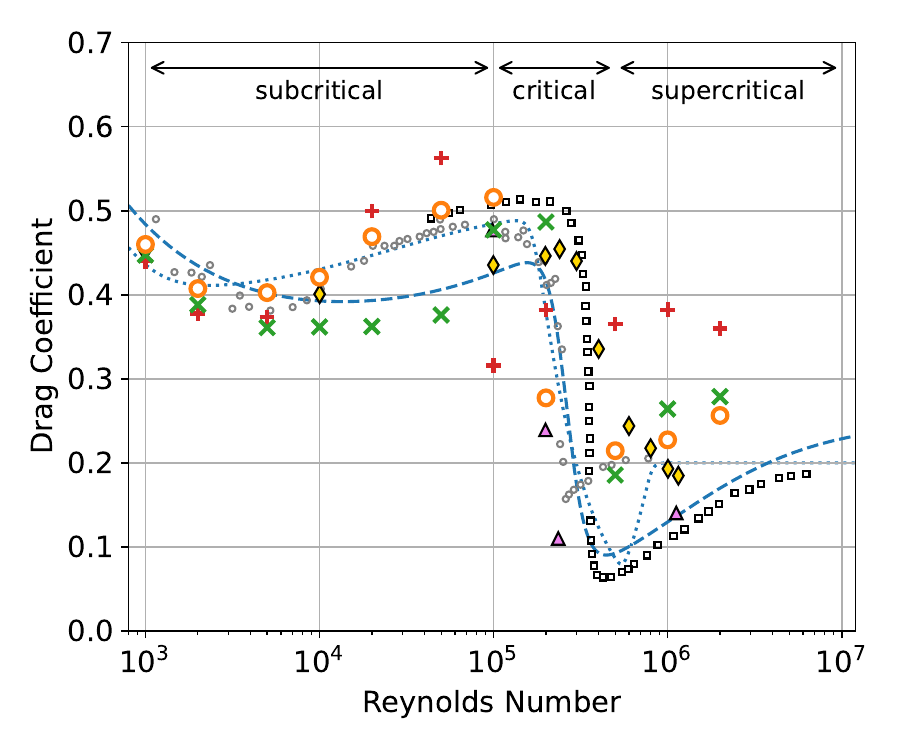}
\subcaption{Close-up of the drag crisis.}
\label{fig:sphere_high_re}
\end{subfigure}
\caption{Drag coefficient for the flow around a sphere for a wide range of Reynolds numbers. The results of the boundary conditions PSM, QBB and no-slip are compared to reference data from  \cite{almedeijDragCoefficientFlow2008}, \cite{morrisonDataCorrelationDrag2016}, \cite{achenbachExperimentsFlowSpheres1972}, \cite{GDZPPN002503859}, \cite{geierParametrizationCumulantLattice2017a} and \cite{holzerCodeGenerationLattice2025}.}
\label{fig:sphere}
\end{figure*}

\begin{figure*}
     \centering
     \includegraphics[width=0.7\textwidth]{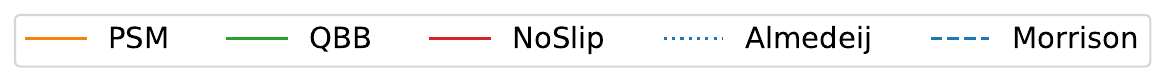}
     \begin{subfigure}[b]{0.45\textwidth}
         \centering
         \includegraphics[width=\textwidth]{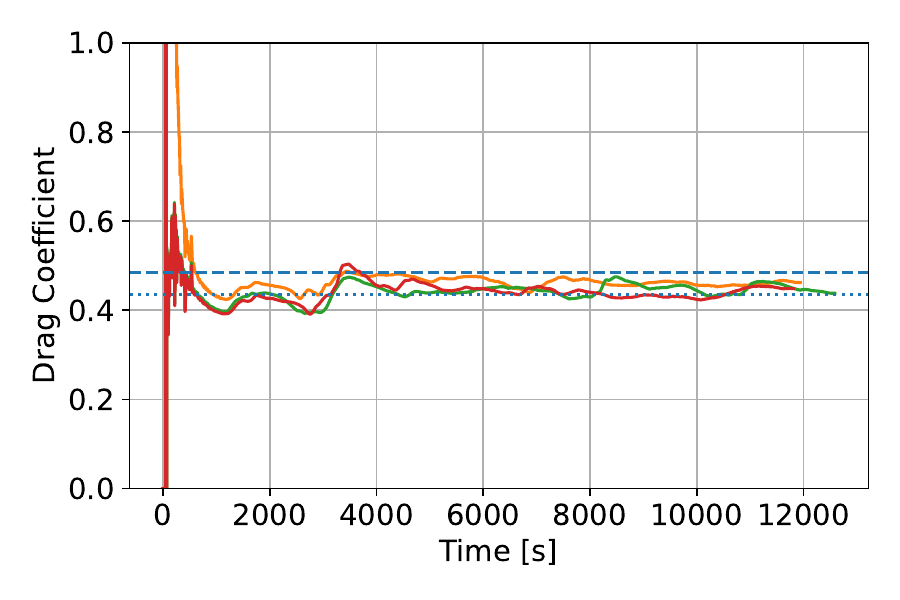}
         \caption{Drag coefficient timeline for Reynolds number 1,000.}
         \label{fig:drag1k}
     \end{subfigure}
     \begin{subfigure}[b]{0.45\textwidth}
         \centering
         \includegraphics[width=\textwidth]{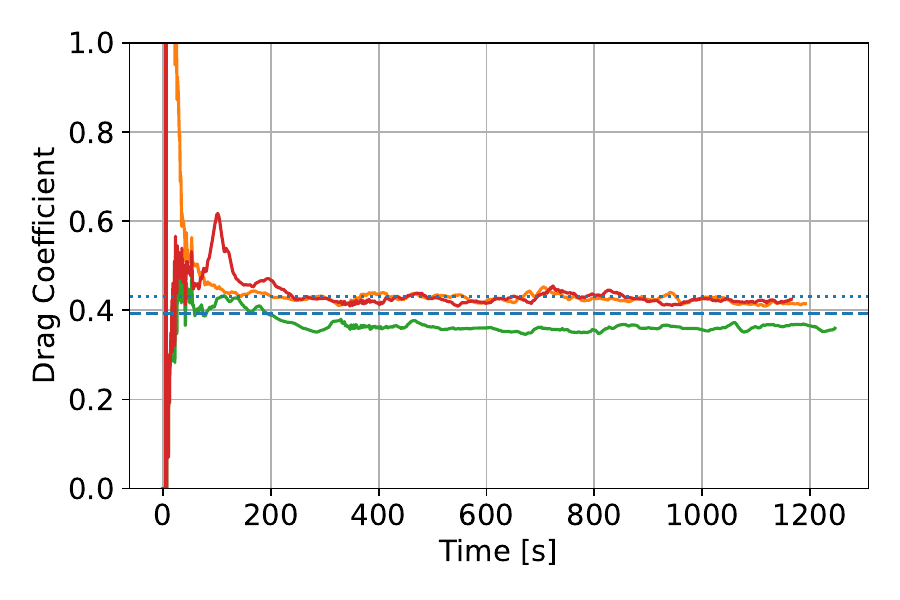}
         \caption{Drag coefficient timeline for Reynolds number 10,000.}
         \label{fig:drag10k}
     \end{subfigure}
     \begin{subfigure}[b]{0.45\textwidth}
         \centering
         \includegraphics[width=\textwidth]{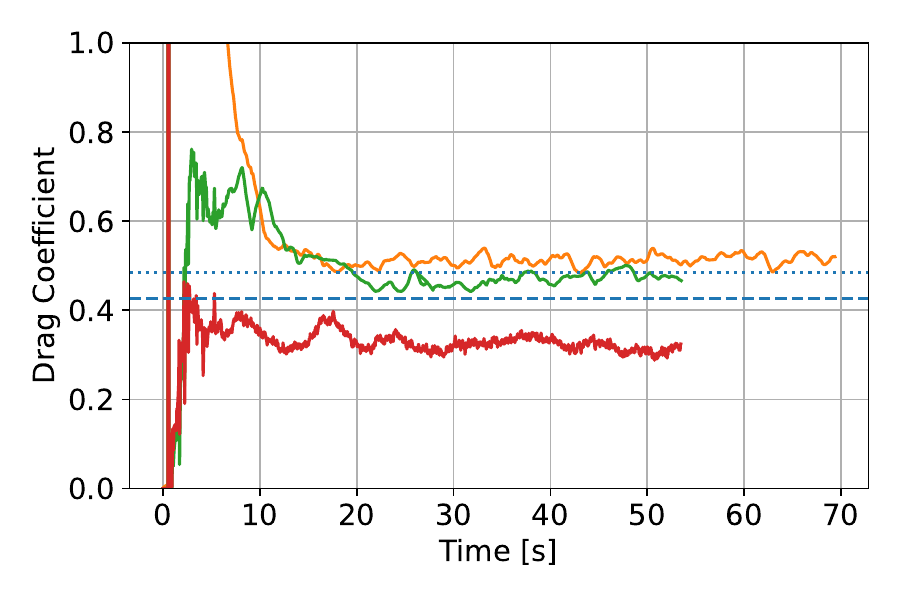}
         \caption{Drag coefficient timeline for Reynolds number 100,000.}
         \label{fig:drag100k}
     \end{subfigure}
     \begin{subfigure}[b]{0.45\textwidth}
         \centering
         \includegraphics[width=\textwidth]{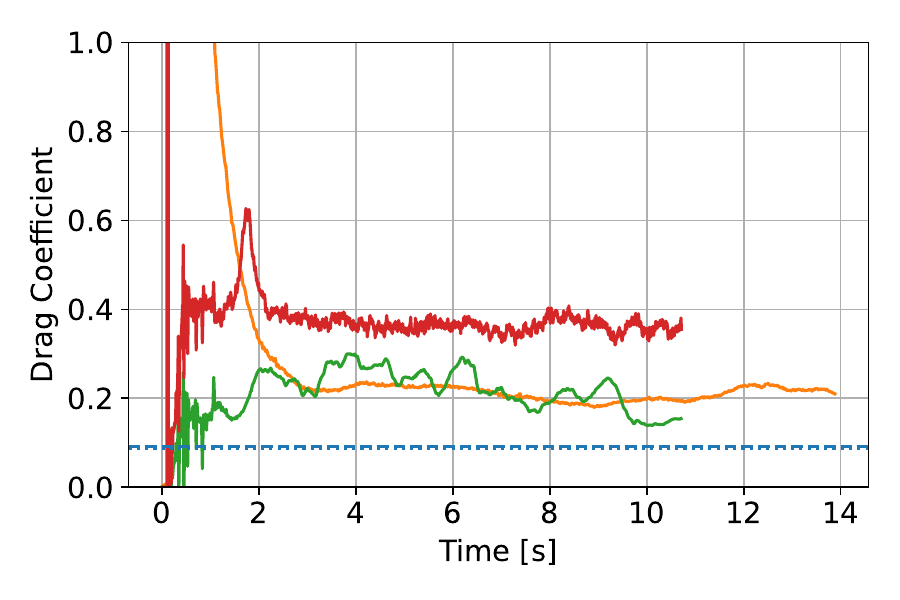}
         \caption{Drag coefficient timeline for Reynolds number 500,000.}
         \label{fig:drag500k}
     \end{subfigure}
        \caption{Comparison of the boundary conditions PSM, QBB and no-slip in terms of the drag coefficient over time for the flow around sphere test case.}
        \label{fig:dragOverTime}
\end{figure*}

\subsection{Results for lower Reynolds numbers}

In this section, we discuss the results of the flow around sphere validation case for the lower Reynolds numbers, i.e., $Re < 10^3$.
In \autoref{fig:sphere_low_re}, we show a comparison of three study cases: 
We realize the boundary of the sphere with three boundary conditions, the PSM, a no-slip and a QBB boundary. 
We compare the time-averaged drag coefficient with reference data from \cite{almedeijDragCoefficientFlow2008} and \cite{morrisonDataCorrelationDrag2016}, and achieve a good agreement for all boundary conditions.
For the PSM, the error with respect to the reference data from \cite{almedeijDragCoefficientFlow2008} lies below 1\% for all runs with $Re<5$ and below 10\% for $Re < 10^3$.
For the no-slip and QBB boundaries, the error lies below 12\% for all cases with  $Re<10^3$.
To decrease these errors, further fine-tuning would be needed per simulation, for example by finding optimal parameters for the inflow velocity, the relaxation rate or the domain setup, or by testing different collision operators. Because of the wide range of Reynolds numbers covered in this study, it is not feasible to fine-tune every single simulation.

\subsection{Results for high Reynolds numbers}

The drag coefficient for varying Reynolds numbers for simulations with turbulent behavior ($Re \geq 10^3$) is displayed in \autoref{fig:sphere_high_re}.
Again, we show results for the PSM and compare them to the no-slip and QBB boundaries. 
For reference, we show data from correlation fit in \cite{almedeijDragCoefficientFlow2008} and \cite{morrisonDataCorrelationDrag2016}, experimental data from \cite{GDZPPN002503859} and \cite{achenbachExperimentsFlowSpheres1972}, and LBM simulation data from \cite{holzerCodeGenerationLattice2025} and \cite{geierParametrizationCumulantLattice2017a}.

In \autoref{fig:sphere_high_re}, the behavior of the drag crisis is clearly visible.
While the drag coefficient is relatively constant for Reynolds numbers between $10^3$ and $10^5$ (subcritical zone), it drops at its trough to around $20\%$ of the original drag value at $Re \sim 5\cdot 10^5$ (critical zone) and recovers to around $40\%$ of the original value for $Re \geq 10^6$ (supercritical zone).

It is evident that the exact drag coefficient values from the references do not fully match each other, especially at the trough.
Nevertheless, the reference data shows a trend for the drag coefficients that our simulations should follow.

\subsubsection{PSM boundary condition}
The data points from our simulations with the PSM show very good agreement with the reference data for Reynolds numbers between $10^3$ and $10^5$. 
In the trough of the drag crisis at $Re \sim 5\cdot 10^5$, the drag coefficient drops to $\sim 40\%$ of its pre-crisis value, but not further.
This may be a result of under-resolving the boundary layer in these highly turbulent flow conditions.
Nevertheless, the PSM predicts the drag crisis at the right Reynolds number, and is in very good agreement especially with the reference data points from \cite{GDZPPN002503859}.

In \autoref{fig:velFields}, the velocity fields of the flow around the sphere are displayed for $Re=10^3$, $Re=10^4$, $Re=10^5$ and $Re=5 \cdot 10^5$.
Here too, the typical behavior of the drag crisis is observable:
While the boundary layer for the subcritical Reynolds numbers separates at an angle of around $80^\circ$, for the supercritical flow at $Re=5 \cdot 10^5$ the boundary layer separates far behind the equator of the sphere at around $120^\circ$.
The wake width also shows its characteristic behavior for supercritical flows in \autoref{fig:vel500k}; it is much narrower than the wake for subcritical Reynolds numbers, as for example in \autoref{fig:vel100k}.

In \autoref{fig:dragOverTime}, we display the drag coefficient over the simulation runtime for the Reynolds numbers $10^3$, $10^4$, $10^5$ and $5 \cdot 10^5$.
As reference, we again show the correlation between Reynolds number and drag coefficient of \cite{almedeijDragCoefficientFlow2008} and \cite{morrisonDataCorrelationDrag2016}.

It is evident that the force signal of the PSM is much more stable. 
This is because the force of the PSM is calculated over the volume of the sphere 
in \autoref{equ:forcing}, while the forces on the no-slip and QBB boundaries are calculated on the surface of the sphere, as shown in \autoref{equ:qbb_force}.
On the other hand, this leads to a slower change of the forces for the PSM, because the forces on the surface of the sphere 
must propagate through the whole 
volume of the sphere.
This can lead to a slower convergence of the drag force for the PSM,
as visible for example in \autoref{fig:drag500k}, where the drag coefficient needs around three seconds to recover from the initially high value that arises when the flow first hits the spherical obstacle.

In \autoref{fig:sphere_high_re}, we also display the drag coefficients for the no-slip and QBB boundary for comparison to the PSM. 
For the no-slip and QBB boundaries in the simulations with $Re \geq 10^5$, we use the Cumulant collision operator with a fourth-order correction and a limiter of 0.01, similar to \cite{holzerCodeGenerationLattice2025} and \cite{geierParametrizationCumulantLattice2017a}.
Without the fourth-order correction, the drag crisis does not arise for these boundary conditions, as shown in \cite{geierParametrizationCumulantLattice2017a}.
For the PSM, no fourth-order correction is needed to develop the drag crisis.

\subsubsection{QBB boundary condition}
The drag coefficients for the QBB boundary for $Re > 10^5$ are at the lower end of the reference data, following the correlation from \cite{morrisonDataCorrelationDrag2016}.
At $Re = 10^5$ there is a jump in the drag coefficient, which may appear because of the change in the resolution of the sphere diameter from 250 cells to 1000 cells (see \autoref{tab:domain}).
This may be a sign of under-resolving the boundary layer of the sphere.
Such a jump in the coefficient when changing the mesh resolution is not observed for the PSM.
To make the boundaries comparable and to save computational resources, we keep the resolution of 250 cells for the Reynolds numbers from $10^3$ to $10^4$.
In the critical zone of the drag crisis, the QBB boundary also follows the reference data from \cite{GDZPPN002503859}.
In contrast to the PSM, the drag coefficient for $Re=2 \cdot 10^5$ does not drop yet, but for $Re=5 \cdot 10^5$, the drag coefficient also drops to $\sim 40\%$ of its pre-crisis value.
When looking at the drag coefficients over time in \autoref{fig:dragOverTime}, we observe a steady force signal for the Reynolds numbers $10^3$ and $10^4$.
This is different for $Re= 5 \cdot 10^5$, where the drag coefficient shows a large variation over time.
This unsteady signal of the QBB boundary is also reported in \cite{geierParametrizationCumulantLattice2017a} and \cite{holzerCodeGenerationLattice2025}.

\subsubsection{No-slip boundary condition}
For the no-slip boundary condition, the drag coefficients in \autoref{fig:sphere_high_re} show a different behavior.
While for $Re \leq 10^4$ the drag is still aligned with the reference data, for higher Reynolds numbers the boundary condition does not capture the physical effects of the drag crisis well.
It predicts the critical region of the crisis earlier, at $1 \cdot 10^5$, and only drops to a drag coefficient of $\sim 0.3$.
In \autoref{fig:dragOverTime}, the force signal of the no-slip boundary condition is very noisy for the cases with $Re \geq 10^5$.
This may be a result of the 
staircase approximation of the curved surface that is characteristic of the no-slip boundary condition. 
We conclude that the first-order 
no-slip boundary condition is 
insufficient to reproduce the drag crisis for this simulation setup correctly.

\subsection{Summary}

Overall, the PSM boundary condition shows good agreement with the literature for the drag coefficient for all Reynolds numbers, comparable to the second-order accurate QBB boundary condition from \cite{geierParametrizationCumulantLattice2017a}.
For very high Reynolds numbers in the critical and supercritical zones of the drag crisis, the problem is not fully resolved.
We already use a high resolution of the problem with 1000 cells for the diameter of the sphere, a problem size of $2.62 \cdot 10^9$ cells and a computational effort of over 3000 GPU hours per simulation;
a further increase of the resolution is not possible with the available computational resources.
Nevertheless, to increase the accuracy of the method for the drag crisis, either further refinement to fully resolve the boundary layer or a suitable wall model is needed.

%% file: chapters/airfoil.tex
\section{Stall prediction for Airfoil NACA0018}

Before turning to moving geometries, we apply the PSM to a second stationary configuration of high practical relevance: the flow around an airfoil, for which we assess the prediction of stall for the NACA0018 profile.

Airfoil-resolved CFD simulations are essential for understanding and predicting the aerodynamic behavior of lifting surfaces under realistic operating conditions.
Unlike simplified models, high-fidelity CFD resolves pressure distributions, boundary-layer development, separation, and stall mechanisms with a spatial and temporal detail that cannot be captured experimentally at comparable cost.
This enables the systematic evaluation of design variations, operating points, and flow-control strategies, and makes airfoil CFD a critical tool for the development of modern aerodynamic components such as wind-turbine blades, aircraft wings, and turbomachinery profiles.

The angle of attack denotes the angle between the chord line of the airfoil and the direction of the oncoming flow.
With increasing angle of attack, the lift of an airfoil first grows nearly linearly until, beyond a critical angle, the flow separates from the suction side and the lift drops abruptly while the drag rises sharply -- a phenomenon known as stall.
When this occurs under steady inflow at a fixed angle of attack, it is referred to as static stall.
Its accurate prediction is a demanding benchmark for any numerical method, as it requires resolving laminar boundary-layer separation,
transition in the separated shear layer, and, at pre-stall angles of attack, turbulent reattachment into a laminar separation bubble.
Reference data from scale-resolving simulations exists mainly for symmetric profiles at moderate Reynolds numbers:
DNS studies of the National Advisory Committee for Aeronautics (NACA) 0012 airfoil profile at $Re = 5\cdot10^4$ resolved laminar separation bubbles and transition (\cite{jonesDirectNumericalSimulations2008}) as well as the massive suction-side separation in full stall (\cite{rodriguezDirectNumericalSimulation2013}), and recently a first DNS dataset for the NACA0018 itself was presented for $Re = 30{,}000$ (\cite{rogowskiDNSNACA00182025}).
At higher Reynolds numbers, a wall-resolved LES of an airfoil near stall at $Re = 2.1\cdot10^6$ was conducted by \cite{asadaWallresolvedLES2018}, illustrating the enormous computational cost of scale-resolving airfoil simulations.
Fully three-dimensional, scale-resolving studies in the intermediate and high turbulent regime around and above $Re \sim 10^5$ therefore remain scarce.

RANS-based approaches, in contrast, are computationally affordable but systematically fail at stall: common one- and two-equation models over-predict the static stall angle by around $2^\circ$ or miss stall entirely (\cite{nivedRANSStallAssessment2022}).
Hybrid RANS-LES methods improve the prediction of massively separated post-stall flow (\cite{spalartDetachedEddySimulation2009}), which has also been demonstrated for the NACA0018 (\cite{guoNACA0018DES2022}), but they retain known weaknesses for the shallow separation bubbles and thickening boundary layers that govern the stall onset (\cite{spalartDetachedEddySimulation2009}).

For the lattice Boltzmann method, airfoil simulations have mostly been restricted to two dimensions or to modeled turbulence, for example 2D MRT-LBM simulations of the NACA0012 at $Re = 10^5$ coupled with a Spalart--Allmaras turbulence model (\cite{chenLBMTurbulentAirfoil2012}), a Spalart--Allmaras-based multi-domain LBM for turbulent airfoil flows (\cite{pellerinSpalartAllmarasLBM2015}), and wall-modeled LBM--RANS simulations of high-Reynolds-number airfoil flow (\cite{wilhelmWallModelLBM2018}).
Highly resolved, three-dimensional LBM simulations of airfoil stall are rare in literature. 
In \cite{ribeiroLatticeBoltzmannSimulationsOscillating2016}, a three dimensional oscillating NACA0012 airfoil geometry was studied, in \cite{konigLatticeBoltzmannAnalysisThreeDimensional2015}, the influence of three-dimensional ice-shapes on the airfoil properties was evaluated, and in \cite{Kutscher2021}, porous trailing edges were studied with a quasi-2D airfoil simulation.

In the following, we validate our simulation results against the wind-tunnel measurements of \cite{timmerNACA00182008} and the experimental and numerical results of \cite{bianchiniExperimentalNumericalAssessment2016}, which provide lift and drag polars for the NACA0018 at a Reynolds number of $1.5\cdot10^5$.
An annotated collection of further NACA0018 reference data is given in \cite{melaniAnnotatedDatabase2019}.

\begin{figure*}
\centering
\includegraphics[width=0.7\textwidth,]{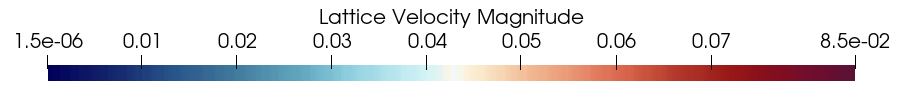}
\begin{subfigure}[b]{0.49\textwidth}
\includegraphics[width=\linewidth]{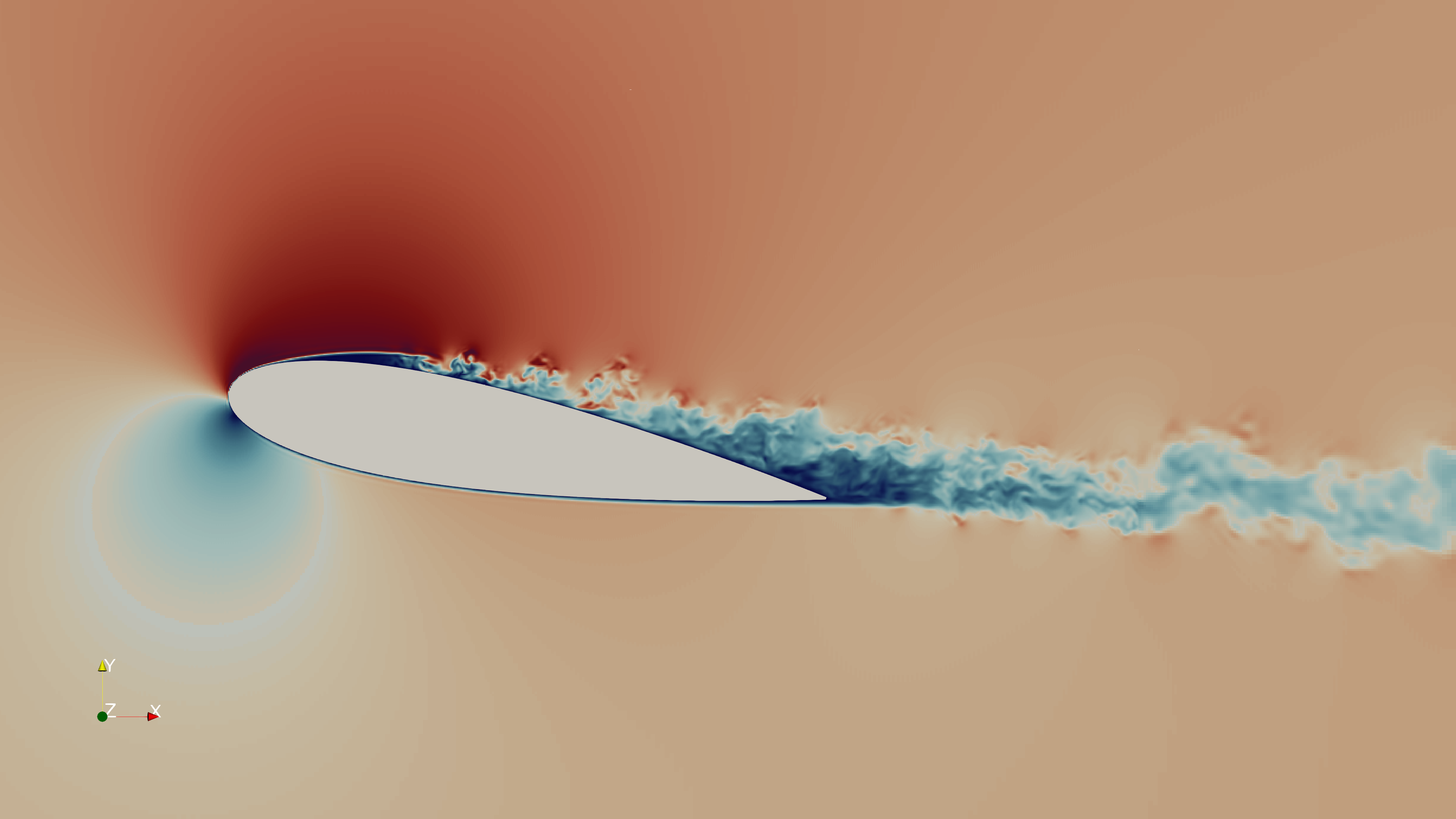}
\subcaption{Pre-stall angle of attack of $10^\circ$.}
\label{fig:aoa_10}
\end{subfigure}
\begin{subfigure}[b]{0.49\textwidth}
\includegraphics[width=\linewidth]{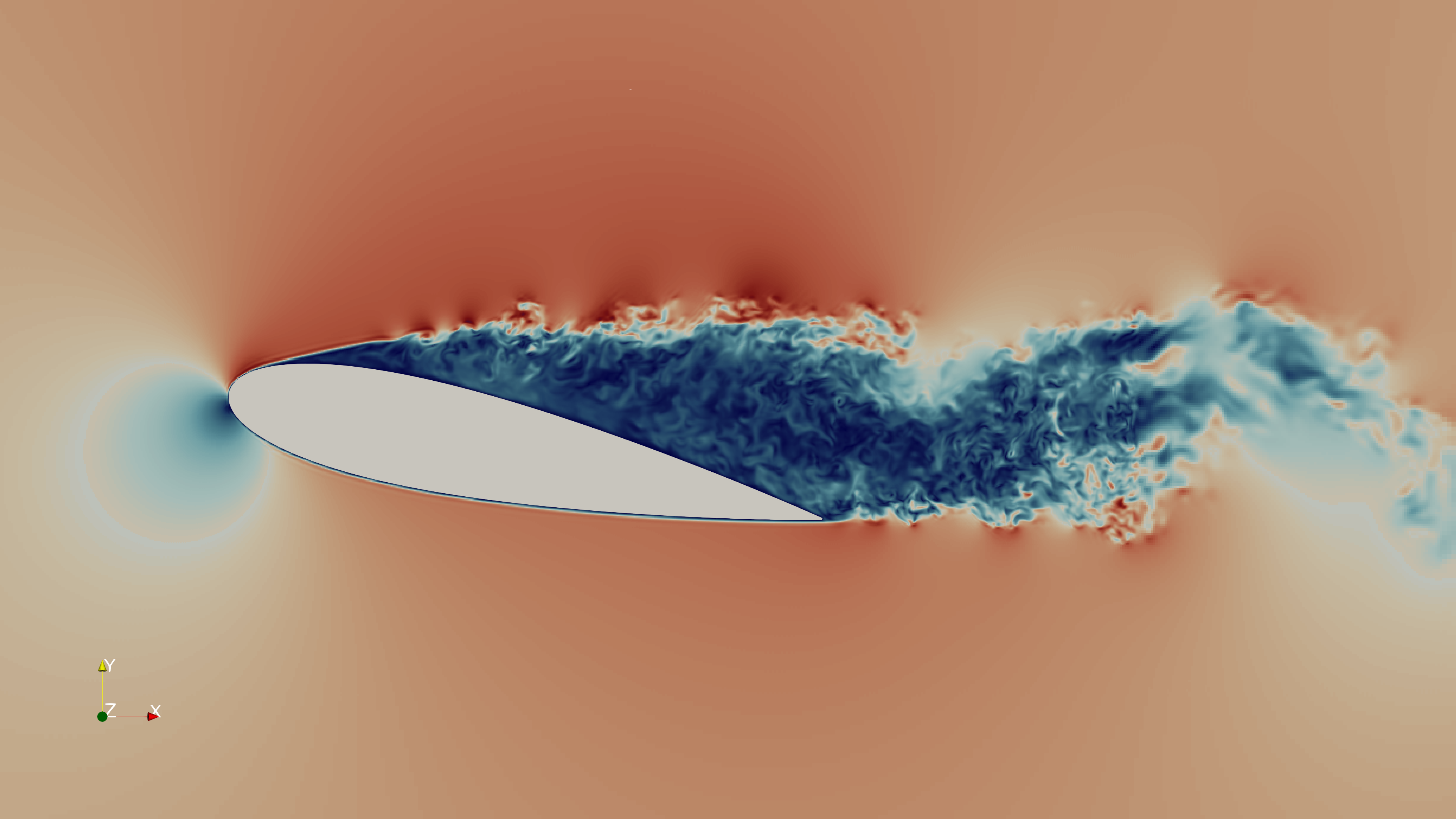}
\subcaption{Stall angle of attack of $12^\circ$.}
\label{fig:aoa_12}
\end{subfigure}
\caption{Velocity magnitude around the NACA0018 airfoil at a pre-stall and at the stall angle of attack.}
\label{fig:airfoil_vel}
\end{figure*}

\subsection{Simulation setup}

The test case considered here is the flow around a NACA0018 airfoil with a chord length of $l_c = 1$ m at a Reynolds number of $1.5 \cdot 10^5$.
The simulation domain measures $20\text{ m} \times 20\text{ m} \times
0.96\text{ m}$, with the airfoil placed at a distance of 7 m from the inflow
boundary.

As for the flow around the sphere in the turbulent regime, the nondimensional inflow velocity
is set to $u_0 = 0.05$ in lattice units, and the relaxation rate $\omega$ is
derived from the target Reynolds number. Again, the Cumulant collision operator with a D3Q27 stencil is employed.
Similar to \autoref{fig:domain}, a velocity bounce-back boundary is used
at the inflow in the western direction and an extrapolation outflow boundary
(\cite{GEIER2015507}) in the eastern direction, while the boundaries normal to the
lift direction are handled by a free-slip condition.
The airfoil spans the full extent of the domain in the spanwise direction, where
periodic boundary conditions are applied.

The spanwise extent of the domain therefore directly determines the largest
turbulent structures that can be represented, which makes it a sensitive parameter for scale-resolving airfoil simulations.
The values reported in the literature vary considerably:
\cite{jonesDirectNumericalSimulations2008} and
\cite{rodriguezDirectNumericalSimulation2013} use a span of 20\% of the chord
length for a NACA0012 at $Re = 5 \cdot 10^4$,
\cite{rogowskiDNSNACA00182025} uses 15\% for the NACA0018 at
$Re = 3 \cdot 10^4$, \cite{asadaWallresolvedLES2018} uses only 5\% at
$Re = 2.1 \cdot 10^6$, and \cite{guoNACA0018DES2022} uses 65\% at
$Re = 10^6$.
More recent studies, however, indicate that such short spans are not always
sufficient.
\cite{turnerKimSpanwiseDNS2020} compare spans of 20\%, 60\%, and 100\% of the
chord length for a stalled airfoil at $Re = 5 \cdot 10^4$ and demonstrate that
larger spans are necessary to predict the radiated noise.
\cite{montalaSpanwiseLength2026} investigate spanwise lengths from 20\% to
200\% of the chord length at $Re = 6 \cdot 10^4$; based on two-point
correlations and flow statistics, they conclude that a minimum spanwise length
of $l_z = 80\%$ of the chord length is necessary to capture the largest
turbulent scales, to reduce the influence of the periodic boundary conditions,
and to obtain accurate predictions of the unsteady aerodynamic forces.
Following these findings, we choose a spanwise length of 96\% of the chord
length to keep the error resulting from an insufficient spanwise
domain small.

The mesh is refined in four levels towards the airfoil, also resolving the flow separation area above the airfoil with the finest mesh level. 
We achieve a resolution of 512 cells per chord length, i.e., a cell size of
$\Delta x \approx 0.002$ m at the airfoil surface, and a total problem size of
$1.7 \cdot 10^8$ cells.
The wall-normal resolution of this mesh is characterized by the dimensionless
wall distance $y^+$ of the first fluid cell layer, which reaches a mean value
of $5.22$ and a maximum value of $25.82$ over the airfoil surface.
The maximum values occur exclusively in cells that are located diagonally to
the boundary, whereas all direct neighbors of the airfoil surface stay below
$y^+ = 10$.

In the pre-stall regime, so for angles of attack $\alpha \leq 10^\circ$, the simulations are run for three flow-through times, that is, three times the time the flow needs to travel from the inflow to the outflow boundary.
For post-stall angles of attack, the flow is considerably more unsteady
and the averaging of the aerodynamic coefficients requires a longer time
series, so these simulations are run for ten flow-through times.
One post-stall simulation consumes around 250 GPU hours on \nvidia H100 GPUs.

For this study, the angle of attack is varied from $0^\circ$ to $30^\circ$ and thus covers the pre-stall
regime, the static stall angle, and the post-stall regime.
The airfoil boundary is again realized with the PSM, the no-slip, and the QBB boundary
condition, so that the three boundary treatments can be compared on an
otherwise identical setup.

\begin{figure*}
\centering
\begin{subfigure}[b]{0.45\textwidth}
\includegraphics[width=\linewidth]{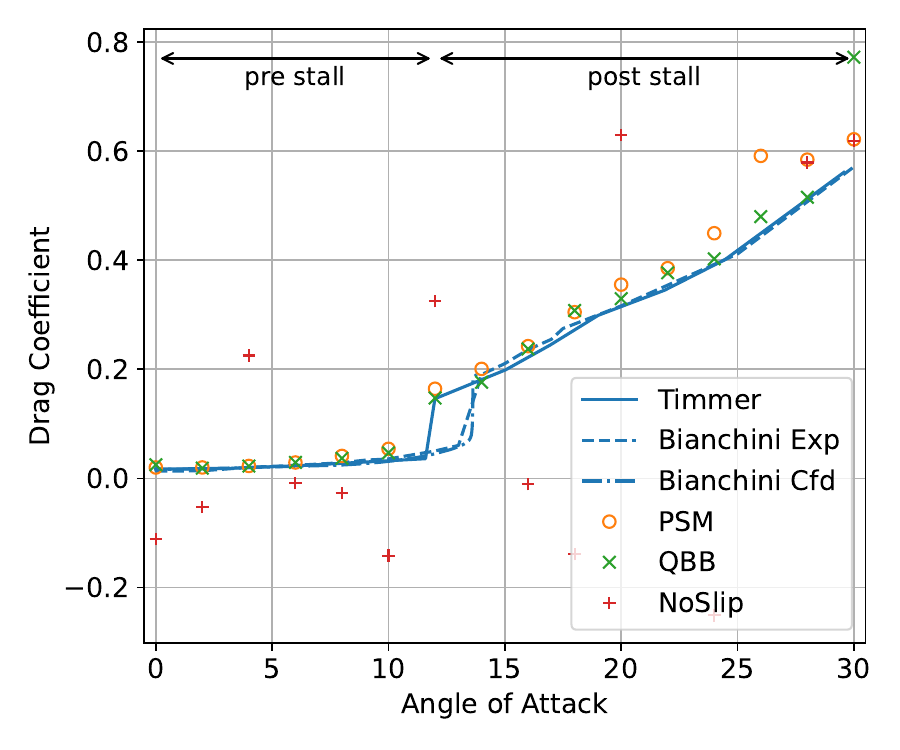}
\subcaption{Drag coefficient.}
\label{fig:airfoil_drag}
\end{subfigure}
\begin{subfigure}[b]{0.45\textwidth}
\includegraphics[width=\linewidth]{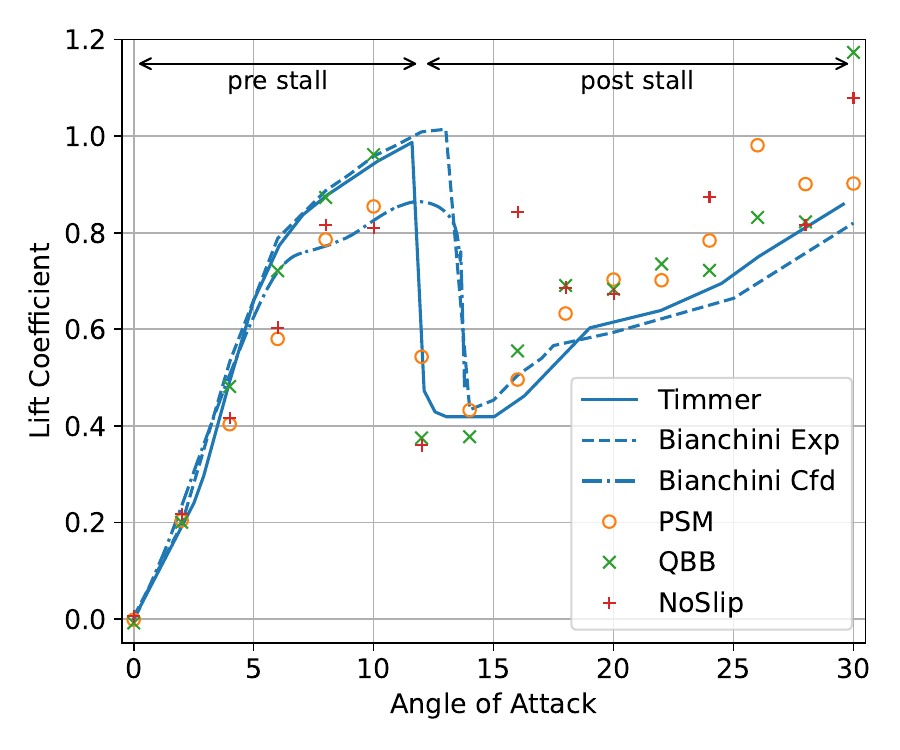}
\subcaption{Lift coefficient.}
\label{fig:airfoil_lift}
\end{subfigure}
\caption{Comparison of the PSM, QBB, and no-slip boundary conditions for the drag and lift coefficient of the NACA0018 airfoil at a Reynolds number of $1.5\cdot10^5$ and varying angles of attack. Reference data from \cite{timmerNACA00182008} and \cite{bianchiniExperimentalNumericalAssessment2016}.}
\label{fig:airfoil_coeffs}
\end{figure*}

\begin{figure}
\centering
\includegraphics[width=\linewidth]{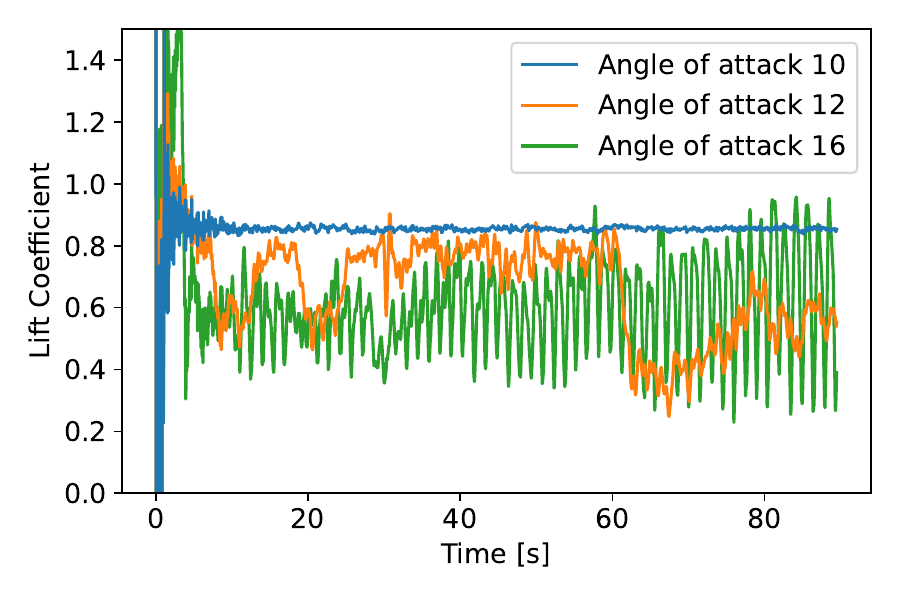}
\caption{Lift coefficient over time for the NACA0018 airfoil resolved by the PSM, for angles of attack of $10^\circ$, $12^\circ$, and $16^\circ$.}
\label{fig:airfoil_timelines}
\end{figure}

\subsection{Results}

We evaluate the drag and lift coefficients of the airfoil for the different angles of attack.
The drag coefficient is computed as
\begin{equation}
    C_D = \frac{2 F_x}{\rho u_0^2 l_{c} l_{z} },
\end{equation}
with $F_x$ as the force integral over the airfoil geometry in flow direction, $l_c$ as the airfoil chord length, and $l_z$ as the airfoil span, i.e., the domain size in the $z$ dimension.
The lift coefficient is calculated analogously from the force $F_y$ normal to the flow direction, so
\begin{equation}
    C_L = \frac{2 F_y}{\rho u_0^2 l_{c} l_{z} }.
\end{equation}

The drag and lift coefficients, time-averaged over the last 25\% of the simulated time, are presented in \autoref{fig:airfoil_coeffs}, together with the experimental results from \cite{timmerNACA00182008} and the experimental and CFD data from \cite{bianchiniExperimentalNumericalAssessment2016} as reference.

For the drag coefficient in \autoref{fig:airfoil_drag}, we observe very good agreement of the PSM and the QBB results with the reference data.
The stall angle is predicted at $12^\circ$, matching the experimental data from \cite{timmerNACA00182008}.

For the lift coefficient in \autoref{fig:airfoil_lift}, the QBB boundary follows the reference data from \cite{timmerNACA00182008} closely for pre-stall angles, and the stall angle is again predicted correctly.
For post-stall angles, however, the QBB overpredicts the lift coefficient by around 10 to 20\%, depending on the angle.
The PSM boundary condition agrees well with the lift coefficients of the CFD data from \cite{bianchiniExperimentalNumericalAssessment2016} for pre-stall angles, predicts the stall angle as precisely as the QBB boundary, and shows the same overprediction of the lift coefficient for post-stall angles.

The reason for the different flow regimes lies in the flow field itself: as presented in \autoref{fig:airfoil_vel}, the flow around a stalled airfoil is much more complex than at pre-stall angles.
At a pre-stall angle of $10^\circ$ (\autoref{fig:aoa_10}), we observe a very thin boundary-layer separation with only a few separation bubbles, while the flow otherwise stays attached.
At an angle of attack of $12^\circ$ (\autoref{fig:aoa_12}), the flow field looks very different: the separated flow region is much wider without any reattachment, and the flow is overall much more unsteady.

This difference is also visible in the force signals.
\autoref{fig:airfoil_timelines} shows the lift coefficient over time for the PSM boundary condition at angles of attack of $10^\circ$, $12^\circ$, and $16^\circ$.
For the pre-stall angle of $10^\circ$, the force signal is very stable, reflecting the simple flow pattern described above.
At the stall angle of $12^\circ$, the lift coefficient becomes noticeably more unsteady, and for the post-stall angle of $16^\circ$, the signal fluctuates strongly.
This unsteady zigzag behavior is present for all post-stall angles and all boundary conditions considered, and it may explain the lower accuracy of the averaged lift coefficients in \autoref{fig:airfoil_lift}.

The overprediction of the lift coefficient for post-stall angles may be reduced by improving the simulation setup in the future.
We apply the same mesh and geometry resolution for pre- and post-stall angles, but a resolution of 512 cells per chord length with an average $y^+$ of around 5 appears to be insufficient for the post-stall regime.
To further improve the accuracy of these simulations, a direct numerical simulation with $y^+<1$ should be performed, which is not possible within this study due to limited computational resources.

Lastly, the no-slip boundary condition completely fails to predict the drag coefficient for any angle of attack and only shows good results for the lift coefficient at pre-stall angles of attack.
Due to its first-order accuracy, the no-slip boundary condition is not suited to capture the physics of this airfoil simulation with the given setup.

Overall, the PSM as well as the QBB boundary condition show good agreement with the reference data for the drag and lift coefficients of the NACA0018 airfoil, with a moderate overprediction of the lift coefficient remaining for post-stall angles of attack.

%% file: chapters/stirredTank.tex
\section{Simulation of a Stirred Tank}

For the two stationary configurations, the sphere and the NACA0018 airfoil, the PSM showed good agreement with reference data from the literature.
We now turn to a turbulent flow with a moving geometry: a stirred tank, in which the flow is driven by a rotating impeller.

Stirred tanks are among the most widely used process devices in the chemical, pharmaceutical, and biotechnological industries, where they drive operations such as blending, chemical reactions, crystallization, and fermentation (\cite{paulHandbookIndustrialMixing2004}).
The design and scale-up of such reactors depends strongly on local turbulence quantities in the vessel, so that the accurate prediction of the impeller-driven turbulent flow field is of direct industrial relevance.
At the same time, the stirred tank is a demanding numerical benchmark, since the rotating impeller constitutes a complex moving geometry inside a baffled vessel, which body-fitted CFD approaches typically have to address with rotating reference frames or sliding meshes.

For the LBM with its non-conforming Cartesian grid, the stirred tank has therefore been a natural target from early on: the first LBM-based large-eddy simulations of a full baffled vessel were presented in \cite{eggelsDirectLargeeddySimulation1996}, where the action of the revolving impeller on the fluid was modeled by a body-force field varying in space and time, and in \cite{derksenLargeEddySimulations1999}, which refined this idea into an adaptive force-field technique that enforces the no-slip condition at control points on the impeller surface via locally computed body forces.
More recently, GPU acceleration has made transient LBM-LES of stirred vessels affordable for engineering practice: GPU-based LBM-LES coupling the rotating impeller to the fluid with an immersed boundary method reduced the simulation of one impeller revolution to minutes (\cite{shuGPUacceleratedLargeEddy2018}), and LBM-LES has since been validated and adopted as a characterization tool for stirred (bio-)reactors in industrial settings (\cite{kuschelValidationNovelLattice2021, kersebaumSimulatingStirredTank2023}), based on a commercial solver that likewise represents the moving impeller by an immersed boundary approach.
All of these approaches thus couple the rotating geometry to the fluid by forcing-based immersed-boundary-type methods, whereas the suitability of the PSM, which instead enters the LBM through a modified collision operator weighted by the solid volume fraction, has not yet been demonstrated for such configurations.

For the validation of the PSM in this setting, we use the laser-Doppler velocimetry measurements of \cite{wuLaserDopplerMeasurements1989}, who measured mean velocities, turbulence intensities, Reynolds stresses, and energy dissipation rates in a baffled tank stirred by a six-blade Rushton turbine, and whose data has served as an established reference for stirred tank simulations ever since.

For the moving geometries considered in this work, i.e., the stirred tank and the wind turbine, the geometry is represented exclusively by the PSM.
The direct-mapping no-slip and QBB boundary conditions are not applied to these cases, since they would require a dedicated treatment of the fresh-node problem (see \autoref{sec:direct_mapping}), which the PSM avoids entirely.

\subsection{Simulation setup}

\begin{figure}
\centering
\includegraphics[width=0.7\linewidth, trim={6cm 0cm 6cm 0cm},clip]{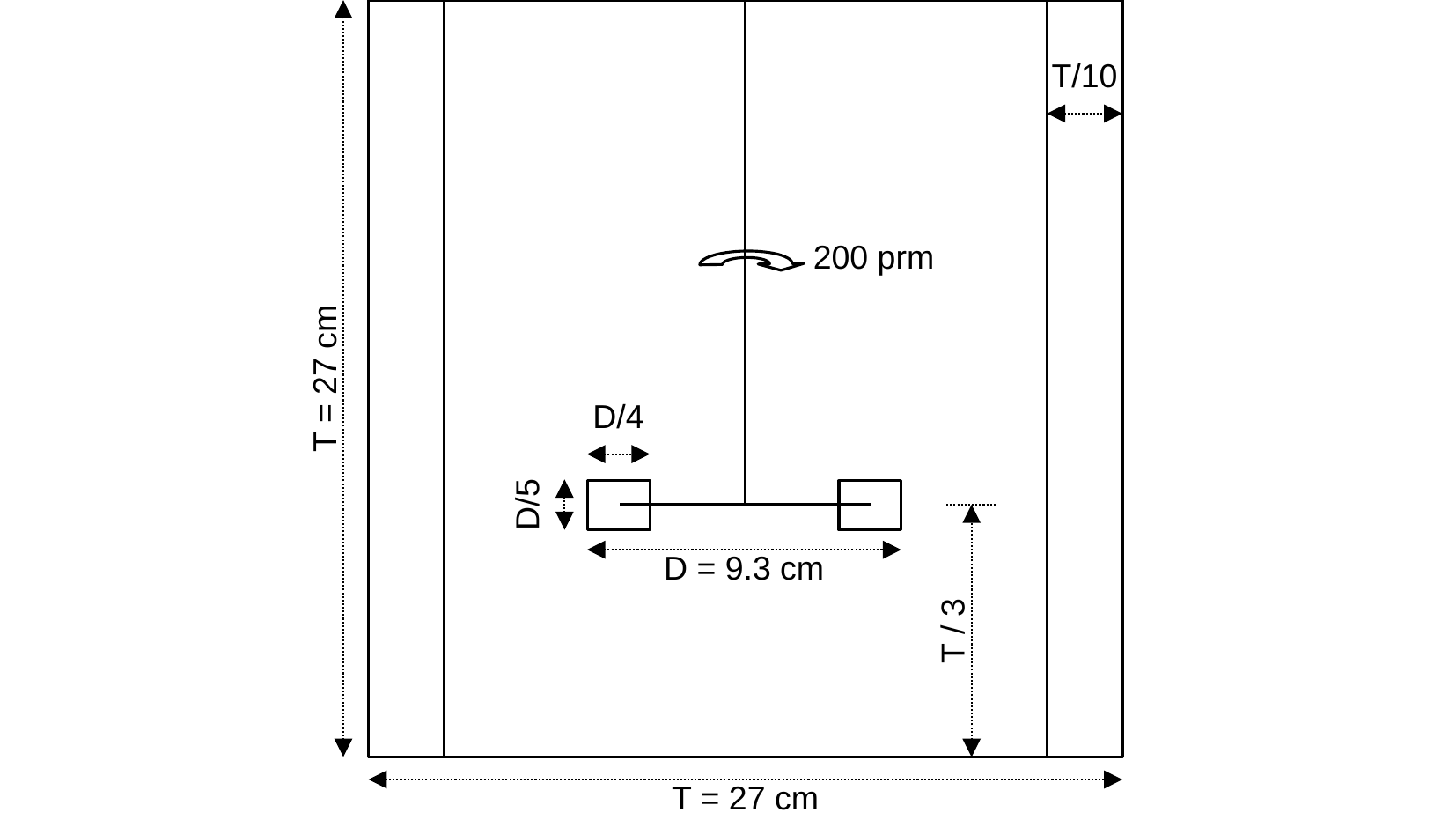}
\caption{Setup for the Stirred Tank simulation.}
\label{fig:stirredTankSetup}
\end{figure}

\begin{figure}
\centering
\includegraphics[width=0.8\linewidth, trim={40cm 0cm 40cm 1cm},clip]{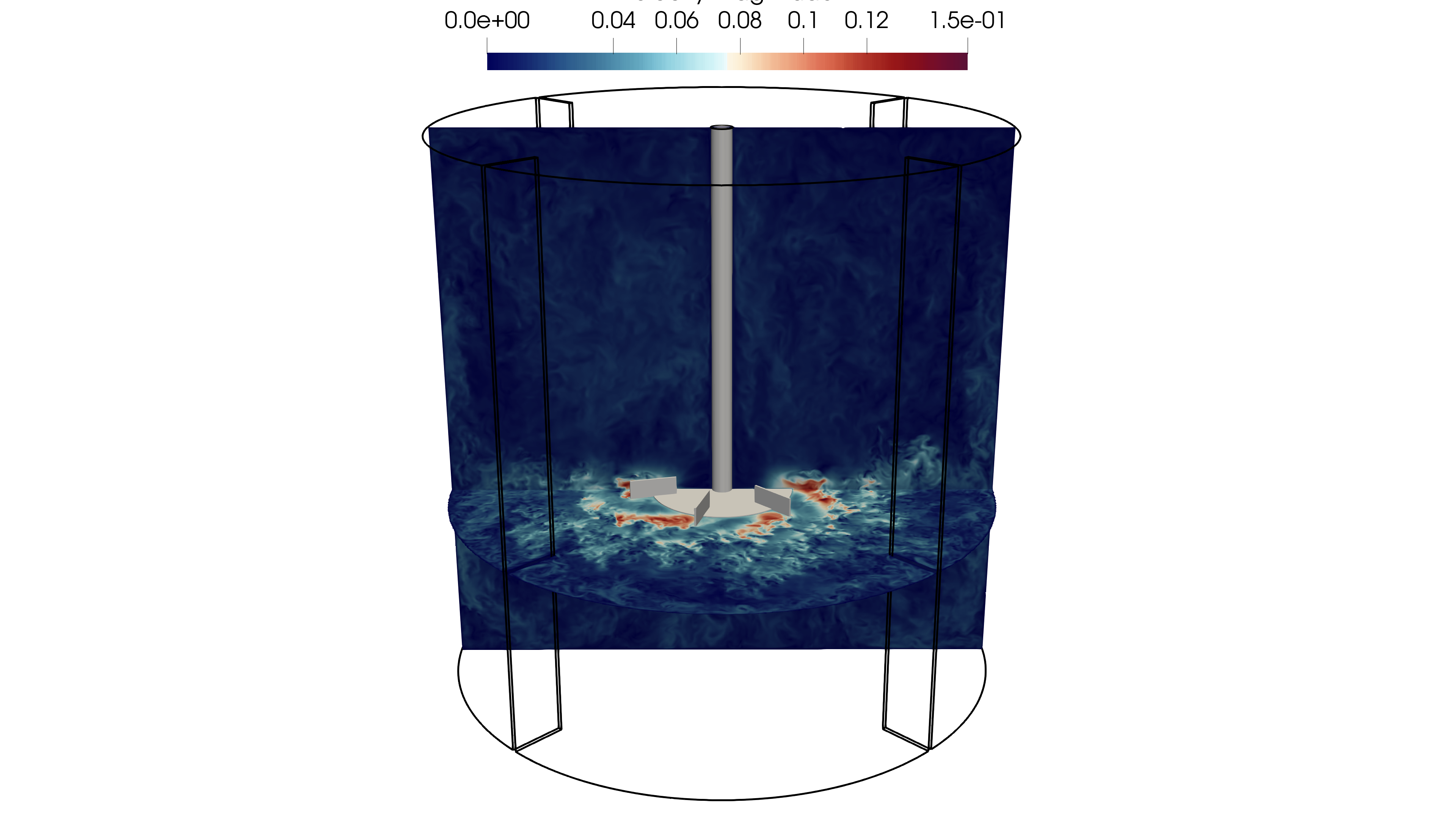}
\caption{Visualization of the stirred tank simulation by the instantaneous velocity magnitude field.}
\label{fig:stirredTankSim}
\end{figure}

The simulation setup mirrors the experimental configuration of \cite{wuLaserDopplerMeasurements1989}, as sketched in \autoref{fig:stirredTankSetup}.
The vessel is a cylindrical tank with a diameter of $T = 27$ cm, equipped with four equally spaced baffles mounted at $90^\circ$ intervals along the tank wall.
The fluid is agitated by a standard six-blade Rushton turbine with a diameter of $D = 9.3$ cm, rotating at a constant speed of $N = 200$ rpm.
The tank is filled with water, so the kinematic viscosity is $\nu = 1 \cdot 10^{-6}$ $\text{m}^2/\text{s}$.
This corresponds to an impeller Reynolds number of
\begin{equation}
    Re = \frac{N D^2}{\nu} = 2.9 \cdot 10^4,
\end{equation}
with the rotation rate $N$ in revolutions per second, so the flow in the vessel is fully turbulent.

The tank diameter is resolved with 700 lattice cells on a uniform Cartesian grid, which results in a total problem size of $4.5 \cdot 10^8$ cells.
As in the previous test cases, the Cumulant collision operator with a D3Q27 stencil is employed.
Both the rotating impeller and the stationary geometry, i.e., the tank wall, the baffles, and the shaft, are represented by the PSM boundary condition.
The rotating impeller is thereby handled without any rotating reference frame or sliding-mesh technique: in every time step, the solid volume fraction of the cells intersected by the impeller is recomputed from its current angular position, and the PSM collision operator accounts for the local wall velocity of the moving blades.

The simulation covers 20 s of physical time, corresponding to roughly 67 impeller revolutions.
The first 10 s are discarded to allow the flow field to develop from the initial state of rest, and all results presented in the following are time-averaged over the remaining 10 s.
One simulation consumes around 53 GPU hours on \nvidia H100 GPUs.

\begin{figure*}
    \begin{subfigure}{\textwidth}
        \includegraphics[width=0.8\linewidth, trim={0cm 0cm 0cm 0cm},clip]{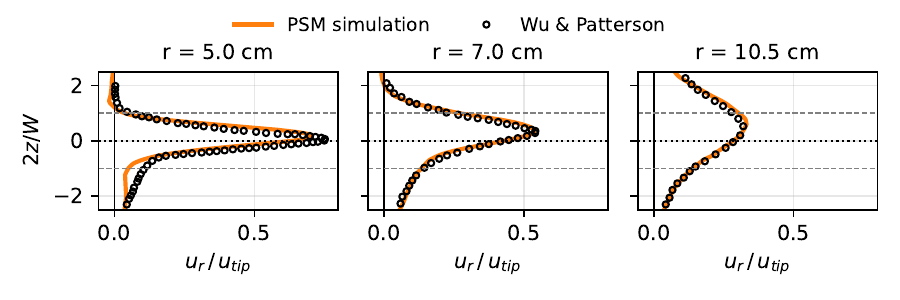}
        \subcaption{Radial mean velocity.}
        \label{fig:tank_radial_vel}
    \end{subfigure}

    \begin{subfigure}{\textwidth}
        \includegraphics[width=0.8\linewidth, trim={0cm 0cm 0cm 0cm},clip]{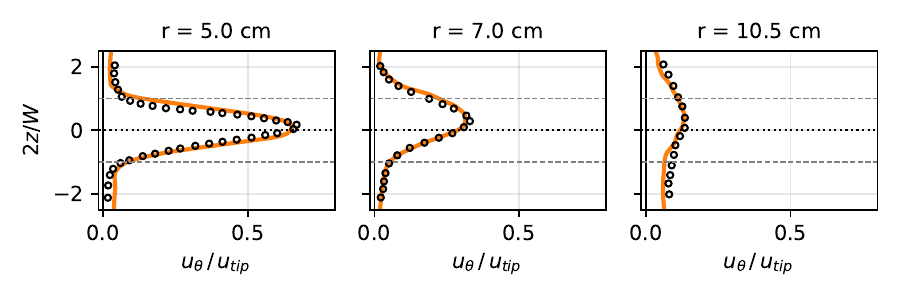}
        \subcaption{Tangential mean velocity.}
        \label{fig:tank_tangential_vel}
    \end{subfigure}

    \begin{subfigure}{\textwidth}
        \includegraphics[width=0.8\linewidth, trim={0cm 0cm 0cm 0cm},clip]{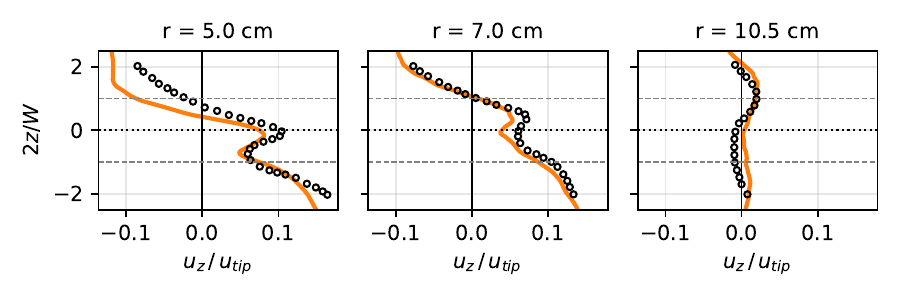}
        \subcaption{Axial mean velocity.}
        \label{fig:tank_axial_vel}
    \end{subfigure}
    \caption{Velocity components for the stirred tank simulation for different distances to the impeller rotation center. Reference data from \cite{wuLaserDopplerMeasurements1989}.}
\end{figure*}

\subsection{Results}
For the validation of our moving geometry approach with the PSM, we compare mean velocity components with the reference data from \cite{wuLaserDopplerMeasurements1989}.
An instantaneous snapshot of the turbulent flow around the impeller is shown in \autoref{fig:stirredTankSim}.
In \autoref{fig:tank_radial_vel}, the radial velocity component $u_r$ normalized with the tip velocity $u_{\text{tip}}$ is presented.
We show data for different distances from the impeller center: a radius of $r = 5$ cm gives information about the velocity components close to the impeller tip, while a radius of $r = 10.5$ cm is further away from the impeller and thus more representative of the bulk flow.
We observe that the normalized radial velocity fits the reference data from \cite{wuLaserDopplerMeasurements1989} very well for all radii.
The tangential and axial velocity components of the PSM simulation in \autoref{fig:tank_tangential_vel} and \autoref{fig:tank_axial_vel} also agree well with the reference data.

To summarize, the PSM approach for the fluid-structure interaction of an impeller geometry accurately reproduces the experimental data for a complex moving geometry in a turbulent flow scenario.

%% file: chapters/turbine.tex
\section{Resolved turbine simulation}
Finally, we apply the PSM to a second turbulent flow around a moving geometry: a blade-resolved wind turbine rotor.

The aerodynamic simulation of wind turbines is a prime example of the complex moving geometries targeted in this work: the rotating blades generate wakes that must be transported over long distances without excessive numerical dissipation, while the flow at the blades themselves governs the loads and the near-wake structure.
Owing to its excellent parallel efficiency on modern GPU-based supercomputers, the LBM has received growing attention in wind energy, as summarized in the recent review of \cite{korbLatticeBoltzmannWindFarm2026}.
Most LBM studies, however, represent the turbine by actuator models, where the blades act on the flow only as body-force distributions:
actuator line simulations with the Cumulant LBM were validated against finite-volume Navier--Stokes results in \cite{asmuthActuatorLineSimulations2020}, an actuator line model implemented in \walberla was shown to match the accuracy of the Navier--Stokes-based solver SOWFA at drastically reduced computational cost (\cite{schottenhammlEvaluationLatticeBoltzmann2022, schottenhammlWalberlawindLatticeBoltzmannbased2024}), and a sliding-mesh actuator line approach was presented in \cite{ribeiroSlidingMeshWindTurbine2024}.
While such models are efficient and accurate for wake and wind-farm studies, they do not resolve the flow at the blades, so that boundary-layer development, flow separation, and the formation of tip vortices at their true scale remain inaccessible.

Geometry-resolved LBM simulations of wind turbines are, in contrast, still scarce.
An early demonstration is given in \cite{deiterdingPredictiveWindTurbine2016}, where an adaptive LBM with a level-set-based embedding of the moving geometry was used to simulate the wakes in an array of three Vestas V27 turbines.
More recently, blade-resolved, wall-modeled LBM simulations of a laboratory-scale rotor were compared to actuator line results and experimental wake measurements in \cite{ribeiroBladeresolvedActuatorLine2025}, and a wall-modeled LES of rotors with the homogenized lattice Boltzmann method, a volumetric solid-fraction coupling closely related to the PSM, was presented in \cite{kummerlanderEfficientWallModelledLarge2025}.

In the following, we simulate a fully geometry-resolved rotating turbine with the PSM and no wall model, using local mesh refinement towards the rotor to achieve a high resolution of the geometry while keeping the computational cost feasible.

\begin{figure*}
    \centering
    \includegraphics[width=\linewidth]{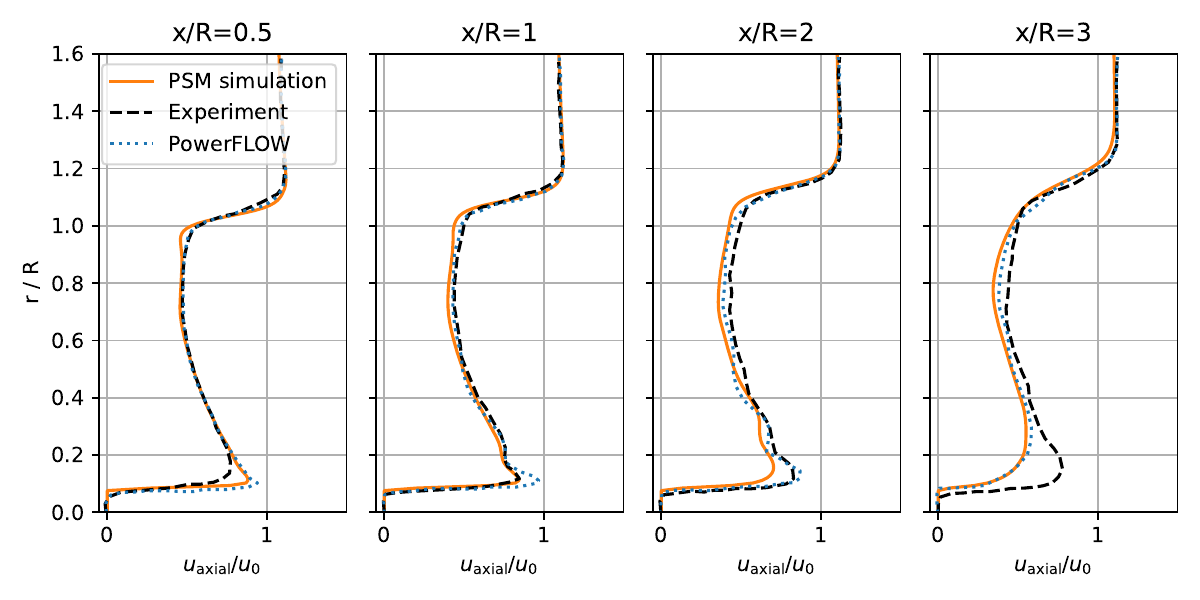}
    \caption{Time and azimuthal-averaged axial velocity at different radial lines. Reference data for experimental and simulation setup from \cite{ribeiroBladeresolvedActuatorLine2025}.}
    \label{fig:turbine_axial}
\end{figure*}

\subsection{Simulation setup}

The simulation setup mirrors the experimental configuration investigated in \cite{ribeiroBladeresolvedActuatorLine2025}, so that the same reference data can be used for validation.
We choose this test case because the rotor geometry is publicly provided by \cite{ribeiroBladeresolvedActuatorLine2025}.
The rotor has a radius of $r = 0.09$ m and a tip chord length of $c_{\text{tip}} = 0.023$ m, and is placed in a domain of $1.2\text{ m} \times 0.5\text{ m} \times 0.5\text{ m}$.
The geometry of the rotor is also presented in \autoref{fig:turbine_pic}.
It rotates at a frequency of $f = 3$ revolutions per second in a fluid with a kinematic viscosity of $\nu = 1.0035 \cdot 10^{-6}\,\text{m}^2\,\text{s}^{-1}$.
A common characteristic number for rotor simulations is the tip-chord Reynolds number
\begin{equation}
    Re_{\text{tip}} = \frac{u_{\text{tip}} c_{\text{tip}}}{\nu}, 
\end{equation}
with the tip velocity $u_{\text{tip}} = 2 \pi r f$.
For the present setup, this yields $Re_{\text{tip}} \approx 3.9 \cdot 10^4$.
The boundary conditions and the sponge zone next to the outflow are chosen similarly to the flow around the sphere setup shown in \autoref{fig:domain}.

As in the previous test cases, we employ the Cumulant collision operator with a D3Q27 stencil.
The rotating rotor and the stationary shaft are both represented by the PSM.
In contrast to \cite{kummerlanderEfficientWallModelledLarge2025}, no wall model is applied, so the flow at the blade surfaces is directly resolved by the boundary treatment.

The mesh is refined in two levels towards the turbine and its near wake up to three rotor radii downstream, since the mean velocity is evaluated in this region in the following.
The refined domain reaches a cell size of $0.0002$ m around the rotor and in the near wake, which corresponds to the resolution used in \cite{ribeiroBladeresolvedActuatorLine2025}, and results in a total problem size of $1.6 \cdot 10^9$ cells.
In \cite{kummerlanderEfficientWallModelledLarge2025}, a uniform grid is applied to the same problem; there, the simulation needs 2.5 times as many cells as our setup for a resolution that is two times coarser.
This again shows the strength of local mesh refinement for turbulent flow problems.

In total, 10 seconds of physical time are simulated, corresponding to 30 rotor revolutions.
The averaging of the results starts after eight warm-up revolutions, i.e., after $2.67$ seconds, so that the initial transient is excluded from the statistics.
One simulation consumes around 125 GPU hours on \nvidia H100 GPUs.

\subsection{Results}

For the validation of our PSM moving geometry approach, we first consider the thrust coefficient, i.e., the dimensionless axial thrust force on the rotor, defined as
\begin{equation}
    C_T = \frac{T}{\frac{1}{2} \rho \pi r^2 u_0^2},
\end{equation}
with $T$ as the integral of all axial forces acting on the rotor.
The reference thrust coefficient from \cite{ribeiroBladeresolvedActuatorLine2025} is $C_T = 0.96$, while we achieve $C_T = 1.00$, corresponding to a deviation of only $4.2\%$.

Further, we evaluate the time- and azimuthally averaged axial velocity along radial lines at different positions behind the rotor. In \autoref{fig:turbine_axial}, we compare our results with the experimental data and the blade-resolved simulation data obtained with the PowerFLOW LBM solver in \cite{ribeiroBladeresolvedActuatorLine2025}.
Excellent agreement between our simulation results and the reference data is achieved.

These results demonstrate that our PSM approach for complex moving geometries is capable of solving turbulent industrial problems such as the simulation of a fully resolved wind turbine.

%% file: chapters/conclusion.tex
\section{Conclusion}

In this study, we presented the capabilities of the Partially Saturated Cells Method for handling the fluid-structure interaction of complex moving objects.
For four test cases in turbulent flow conditions -- the flow around a sphere, an airfoil, a stirred tank, and a wind turbine -- we achieved good agreement with reference data from the literature.
Together with the excellent performance that was already demonstrated in the predecessor of this study (\cite{suffaLargescaleSimulationsFully2025}), this makes the PSM a promising option for simulating complex moving geometries in turbulent flows, as they arise in realistic industrial applications.

For future work, a comparison with other fluid-structure interaction approaches, such as sliding meshes or the immersed boundary method, in terms of accuracy and computational efficiency would be of interest.
However, such a comparison would require considerable effort, since all methods would have to be implemented within the same framework, and it may also be difficult to find a physical problem for which the comparison is fair to all approaches.